\documentclass[preprint, review, authoryear, 12pt, 3p]{elsarticle}

\usepackage[table,xcdraw]{xcolor}
\usepackage{amssymb}
\usepackage{bm}
\usepackage{array}
\newcolumntype{P}[1]{>{\centering\arraybackslash}p{#1}}
\usepackage{threeparttable}  
\usepackage{booktabs}
\usepackage{dirtytalk}
\usepackage{scrextend}
\usepackage[linesnumbered,ruled,vlined]{algorithm2e}
\usepackage{subfig}
\usepackage[hidelinks=true]{hyperref}
\usepackage{hhline}
\usepackage{multirow}
\usepackage{mathtools}
\usepackage{multicol}
\usepackage{amsmath} 

\newcommand{\comm}[1]{}

\usepackage{soul}

\biboptions{sort&compress, semicolon, round}

\journal{Computers in Biology and Medicine}

\begin{document}

\begin{frontmatter}

\title{DSNet: Automatic Dermoscopic Skin Lesion Segmentation}


\author[label2]{Md. Kamrul Hasan\corref{cor1}}
\address[label2]{Computer Vision and Robotics Institute, University of Girona, Spain \fnref{label4}}

\cortext[cor1]{Corresponding author}


\ead{kamruleeekuet@gmail.com}

\author[label2]{Lavsen Dahal}
\ead{er.lavsen@gmail.com }

\author[label3]{Prasad N. Samarakoon}
\ead{prasadnsamarakoon@gmail.com }
\address[label3]{Universit\'e Clermont Auvergne, Clermont-Ferrand, France \fnref{label4}}

\author[label2]{Fakrul Islam Tushar}
\ead{f.i.tushar.eee@gmail.com }

\author[label2]{Robert Mart\'i}
\ead{robert.marti@udg.edu}


\begin{abstract}
\subsection*{Background and Objective}
Automatic segmentation of skin lesions is considered a crucial step in Computer Aided Diagnosis (CAD) for melanoma. Despite its significance, skin lesion segmentation remains an unsolved challenge due to their variability in color, texture, and shapes and indistinguishable boundaries. 

\subsection*{Methods}
Through this study, we present a new and automatic semantic segmentation network for robust skin lesion segmentation named Dermoscopic Skin Network (DSNet). In order to reduce the number of parameters to make the network lightweight, we used a depth-wise separable convolution in lieu of standard convolution to project the learnt discriminating features onto the pixel space at different stages of the encoder. Additionally, we implemented both a U-Net and a Fully Convolutional Network (FCN8s) to compare against the proposed DSNet.

\subsection*{Results}
We evaluate our proposed model on two publicly available datasets, namely ISIC-2017\footnote{\url{https://challenge.kitware.com/\#challenge/583f126bcad3a51cc66c8d9a}} and PH2\footnote{\url{https://www.fc.up.pt/addi/ph2\%20database.html}}. The obtained mean Intersection over Union (mIoU) is $77.5\,\%$ and $87.0\,\%$ respectively for ISIC-2017 and PH2 datasets which outperformed the ISIC-2017 challenge winner by $1.0\,\%$ with respect to mIoU. Our proposed network also outperformed U-Net and FCN8s respectively by $3.6\,\%$ and $6.8\,\%$ with respect to mIoU on the ISIC-2017 dataset.

\subsection*{Conclusion}
Our network for skin lesion segmentation outperforms the other methods discussed in the article and is able to provide better segmented masks on two different test datasets which can lead to better performance in melanoma detection.
Our trained model along with the source code and predicted masks are made publicly available\footnotemark. \\ 

\footnotetext{\url{https://github.com/kamruleee51/Skin-Lesion-Segmentation-Using-Proposed-DSNet} \label{note}} 

\end{abstract}

\begin{keyword}
Skin lesion segmentation \sep Computer Aided Diagnosis (CAD) \sep Melanoma detection \sep Deep Learning.
\end{keyword}

\end{frontmatter}

\section{Introduction}
\label{sec1}

\begin{figure}[!ht]
  \centering \includegraphics[width=16cm, height= 7.5cm]{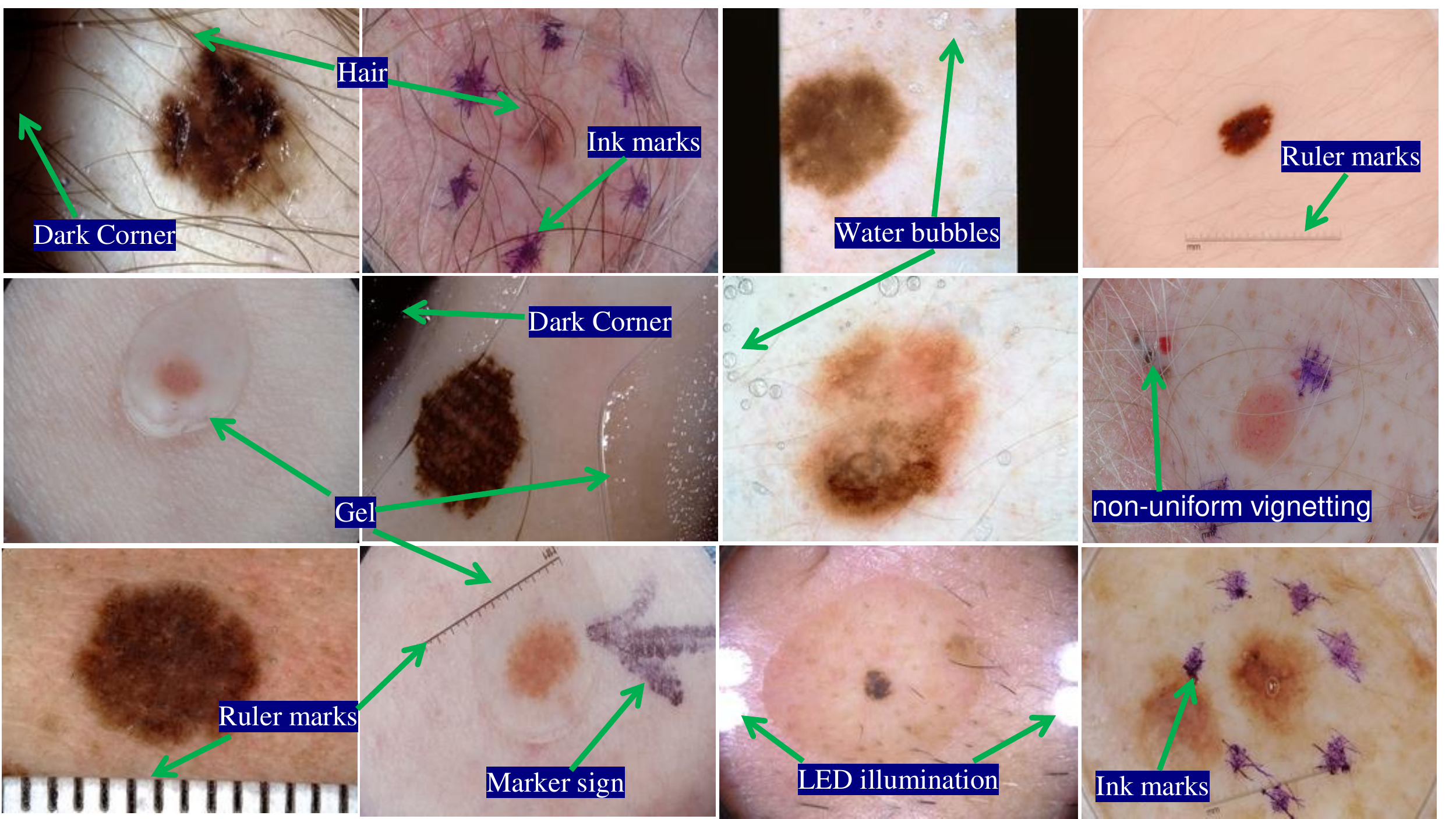}
  \caption{A typical pictorial presentation of the skin images in ISIC-2017 test dataset \citep{ISIC2017data} with different challenging images for segmentation. The obtained segmented masks of these challenging images using our proposed DSNet are provided in \ref{appendix1}.}
  
  \label{fig:TrainingImages}
\end{figure}

One in every three cancers is a skin cancer worldwide \citep{r1}. It is estimated to have 96,480 new cases and 7,230 deaths from melanoma in 2019 in the United States alone \citep{r2}. Melanomas constitute less than $5\,\%$ of all skin cancers, however, account for around $75\,\%$ of all skin-cancer-related deaths in the United States alone \citep{esteva2017dermatologist}. Early diagnosis is very important in the case of skin cancer as shown by the study where the survival rate was as high as $90\,\%$ for melanoma with early detection \citep{r1}. Multiple non-invasive imaging techniques such as dermoscopy, photography, Confocal Scanning Laser Microscopy (CSLM), Optical Coherence Tomography (OCT), ultrasound imaging, Magnetic Resonance Imaging (MRI), and spectroscopic imaging are currently being used to assist dermatologists in skin lesion diagnosis \citep{typeImage}. Traditionally, the images output by the above mentioned techniques are visually inspected by dermatologists to diagnose skin cancer \citep{dermatologist}, which is often considered a complex and tedious task. To assist the dermatologist and to improve the accuracy, Computer Aided Diagnosis (CAD) systems have been developed. Nowadays, CAD has become an essential part of the routine clinical work for abnormality detection in medical images at many screening sites and hospitals \citep{cad}. CAD systems generally consist of multiple components such as raw image acquisition, preprocessing, Region of Interest (ROI) segmentation, responsible feature extraction, and disease classification \citep{cad2,r5}. Segmentation is a vital component of skin cancer diagnosis as the features for the classification are obtained from the ROI of a segmented mask \citep{r28}. However, automatic and robust skin lesion segmentation is a challenging task due to patient specific properties such as skin color, texture, size of lesion area, and the presence of numerous artifacts such as body hair, reflections, air bubbles, rolling lines, shadows, color calibration charts, non-uniform lighting, non-uniform vignetting, markers, and ink \citep{Stoecker} as shown in Fig. \ref{fig:TrainingImages}.

In the early days, a mixture of edge-detection, thresholding, active contours (expectation-maximization, level sets, clustering, adaptive snakes, \textit{etc.}) or region based (region growing, iterative stochastic region merging, \textit{etc.}) techniques were used for skin lesion segmentation \citep{r4, Roberta}. \citet{r7} extracted skin lesion borders by fusing ensembles of several thresholding methods which can usually achieve accurate and robust results for simple textures and when the contrast between the lesion (foreground) and background  is large. Thresholding based methods often fail to obtain a precise segmentation due to the presence of different types of artifacts as mentioned above. A modified region-based active contour detection method was proposed by \citet{Abbas} for multiple lesion segmentation whereas \citet{Zhou} proposed a mean shift based gradient vector flow algorithm to locate the correct borders. However, when the skin lesion border is fuzzy, the above mentioned methods provide only a coarse segmentation. \citet{r5} used saliency based image enhancement followed by Otsu thresholding. Nevertheless, selecting a proper threshold is often unattainable due to noise and intensity variability of the skin images. An automatic segmentation algorithm using color space analysis and clustering-based histogram thresholding was proposed by \citet{r6} to determine the optimal color channel for segmentation of skin lesions. 
This method performed better only when high resolution images were used in addition to requiring preprocessing steps such as a hair removal algorithm. \citet{r11} used median cut, k-means, fuzzy-c means, and mean shift for automatic border extraction of skin lesions. These computer vision algorithms depend highly on parameter tuning, and finding a robust set of parameters that leads to accurate segmentation is often difficult. Recently, semantic segmentation using Convolutional Neural Networks (CNN) has become popular as they lead to robust solutions for object segmentation by pixel wise classification \citep{long, olaf, r12, r13}. \citet{r16} presented a multi-stage Fully Convolutional Network (mFCN) approach for accurate skin lesion segmentation where they employed a parallel integration method to combine the outputs of every stage. 
The first stage consisted of building a probability map using the original image. 
Then, both the original image and the previously built probability map were fed into the remaining (m-1)FCN networks. 
The results produced at different stages of the network were complementary to each other where the early stages were responsible in producing gross skin lesion masks while the later stages were responsible in generating finer lesion boundaries.
\citet{Yujiao} developed a multi-stage UNet (MS-UNet) with deeply supervised learning strategy where they injected multiple U-Nets into the auto-context scheme to segment skin lesions. A full resolution convolutional network (FrCN) was proposed by \citet{Mohammed} where they removed all the sub-sampling layers of the encoder to get the full resolution of the input image without losing any spatial information. However, without sub-sampling, the CNN models often suffer from over-fitting due to redundant features in addition to having a narrower field of view of the feature maps \citep{long}. \citet{r18} introduced SLSDeep, combining skip-connections, dilated residuals, and pyramid pooling networks. In their model, the encoder network depends on dilated residual network layers along with pyramid pooling network with the enhanced ability to learn features from the input dermoscopic images.
The loss function they optimized comprised a Negative Log Likelihood (NLL) term and an End Point Error (EPE) term. \citet{Saleh} proposed DermoNet having an encoder-decoder paradigm where the encoder consisted of multiple dense blocks while the decoder was used to recover the full image resolution. To process high-resolution features from early layers as well as high-semantic features of deeper layers, they linked the output of each dense block with its corresponding decoder. 
\citet{yuan2017automatic} presented a deep fully convolutional-deconvolutional neural network where the learning rate was adjusted based on the first and the second-order moments of the gradient at each iteration. 
In their method, the authors included additional channels such as Hue-Saturation-Value from the  HSV color space and the L channel (lightness) from CIELAB space along with the original RGB channels. 
Moreover, the authors used bagging-type ensemble of six networks to obtain the final lesion mask.

In this article, we propose Dermoscopic Skin Network (DSNet), a semantic segmentation network for automatic and robust dermoscopic skin lesion segmentation. 
To eliminate the necessity of learning the redundant  features in the encoder of the proposed network, we use dense blocks and transition blocks which were inspired by DenseNet \citep{denseNet}. 
In order to reduce the number of parameters that results in a more generic and lightweight network, depth-wise separable convolution has been used in the decoder similar to the studies of \citet{Xception} and \citet{Kaiser}. 
Skip connections \citep{olaf} have also been used to learn back the spatial information lost  due to pooling at each stage of the encoder.
To highlight the importance of the overlapping between the true and predicted skin lesion masks, we propose sum of intersection over union and cross entropy as loss functions. 
The novelty of this paper lies in the methodology where the proposed DSNet has distinct learning capabilities over the existing CNN networks for the skin lesion segmentation while being very light-weight. Transfer learning of CNN by leveraging the knowledge from previously trained model was employed for training the proposed DSNet. 
Our method was able to achieve state-of-the-art results on  both the IEEE International Symposium on Biomedical Imaging (ISBI) and PH2 datasets. 
To the best of our knowledge, the proposed DSNet is the most lightweight network that produces state-of-the-art results.
We also compare the performance of our DSNet against a number of well-known deep learning approaches (e.g. FCN, U-Net, etc.) under the same conditions using the same datasets.


The remaining sections of the article are organized as follows: section \ref{Methodology} describes the architecture of the proposed network and methodologies whereas section \ref{Experiments} describes the experiments and the results obtained along with an interpretation of the results. Finally, section \ref{Conclusions} concludes the article.

\section{Methods and Material}
\label{Methodology}
\subsection{Proposed DSNet Architecture}
In general, CNN for semantic segmentation consists of two essential components: the encoder and the decoder \citep{Vijay}. The encoder is composed of convolution and sub-sampling layers and is responsible for the automatic feature extraction \citep{min}. The convolution layers are used to produce the feature maps whereas the sub-sampling layers are used to achieve spatial invariance by reducing the resolution of these generated maps. This reduction in resolution leads to an expansion of the field of view of the feature map which in turn makes the extraction of more abstract salient features possible in addition to minimizing the computational cost \citep{long}. The decoder semantically projects the discriminating lower resolution features learnt by the encoder onto the pixel space of higher resolution to obtain a dense pixel wise classification \citep{Garcia}. However, the significantly reduced feature maps due to sub-sampling, often suffer from spatial resolution loss which introduces coarseness, less edge information, checkerboard artifacts, and over-segmentation in semantically segmented masks \citep{long, olaf, odena}. To overcome these problems, \citet{olaf} introduced skip connections in a U-Net which allowed the decoder to recover the relevant features learnt at each stage of the encoder that were lost due to pooling. Similarly, \citet{long} fused features at different coarseness levels of the encoder in FCN to refine the segmentation.
However, when the deconvolution kernel size is not divisible by the up-scaling factor, a \textit{deconvolution overlap} occurs as the number of low resolution features that contributs to a single high resolution feature is not constant across the high resolution feature map \citep{odena}. Due to this \textit{deconvolution overlap}, checkerboard artifacts may appear in the segmented mask. In that sense,  \citet{Mohammed} proposed Full resolution Convolution Network (FrCN) which does not have any sub-sampling layers in the encoder to preserve the spatial information of the feature maps for precise segmentation. However, sub-sampling of feature maps is highly desirable to be employed in CNN due to the several positive aspects previously mentioned. All the semantic segmentation networks have similar encoder design but they vary mainly in their decoder mechanism with respect to how the discriminating features are projected onto the pixel space. 

\begin{figure}[!ht]
\centering
\includegraphics[width=15cm,height=5.5cm]{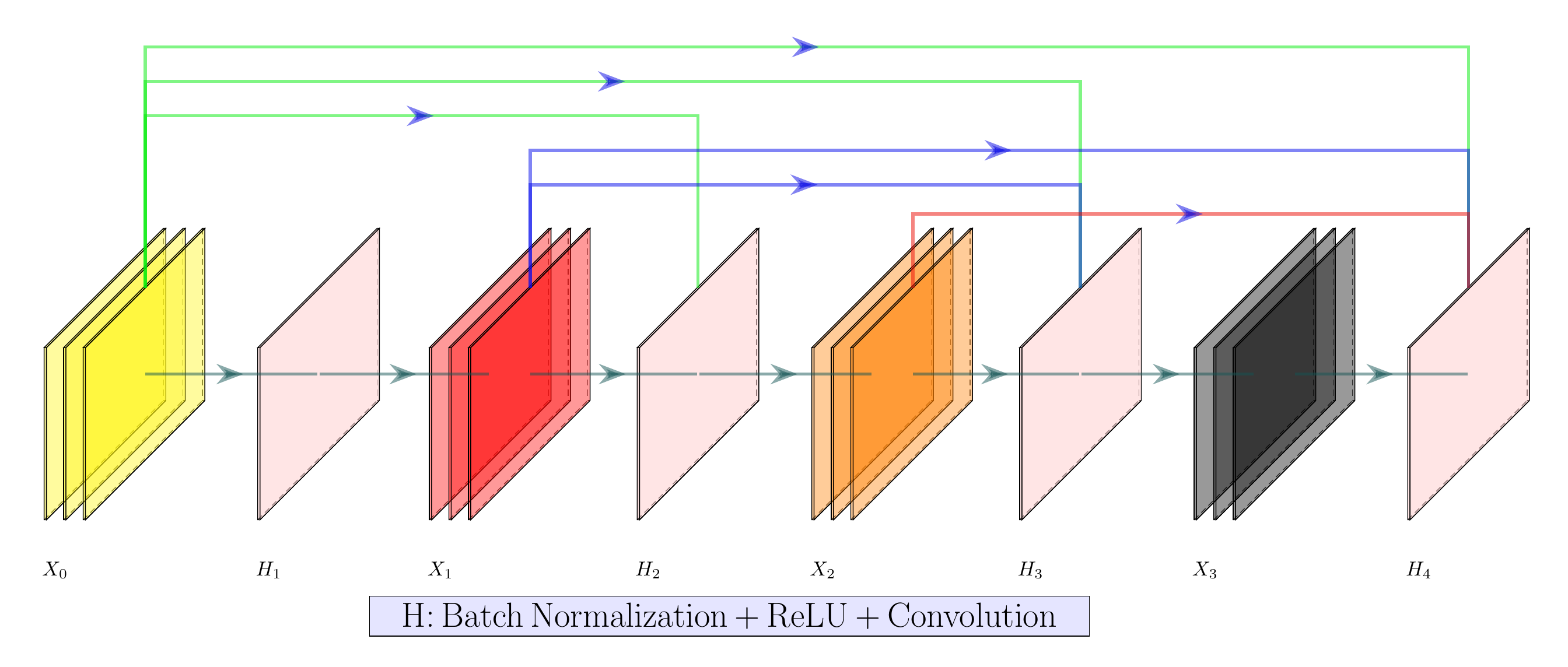}
\caption{A typical example of a dense block of DenseNet with 4 layers and a growth rate of 3 where each feature map $H$ concatenates all the feature maps of its preceding layers of the block for feature reuse throughout the network.  
Consequently, this makes the encoder of DSNet more compact (lightweight).}
\label{fig:encodera}
\end{figure}

In the proposed DSNet, the encoder has $121$ layers which mimics the DenseNet \citep{denseNet} architecture to eliminate the necessity of learning the redundant features. An encoder such as the one proposed in DSNet has the ability to learn the abstract features of lesions from skin images, reduce the vanishing-gradient problem occurrences, and strengthen feature propagation. The design of the DSNet enables each encoding layer to have direct access to the gradients of the loss functions of all previous encoding layers and also to the original input image using skip connections as shown in Fig. \ref{fig:encodera}. 
Each $n^{th}$ layer of a dense block receives the feature maps of all preceding layers which can be formulated \citep{denseNet} as Eq. \ref{eq:feature}. 
\begin{equation} 
\label{eq:feature}
X_n = H_n([X_0, X_1, X_2,...,X_{n-1}])
\end{equation}
\noindent where $[X_0, X_1, X_2,...,X_{n-1}]$ are the concatenation of the feature maps produced in all previous layers and $H_n$ is the composite function \citep{denseNet} of the $n^{th}$ layer. 


\begin{figure}[!ht]
\centering
\includegraphics[width=14cm,height=6cm]{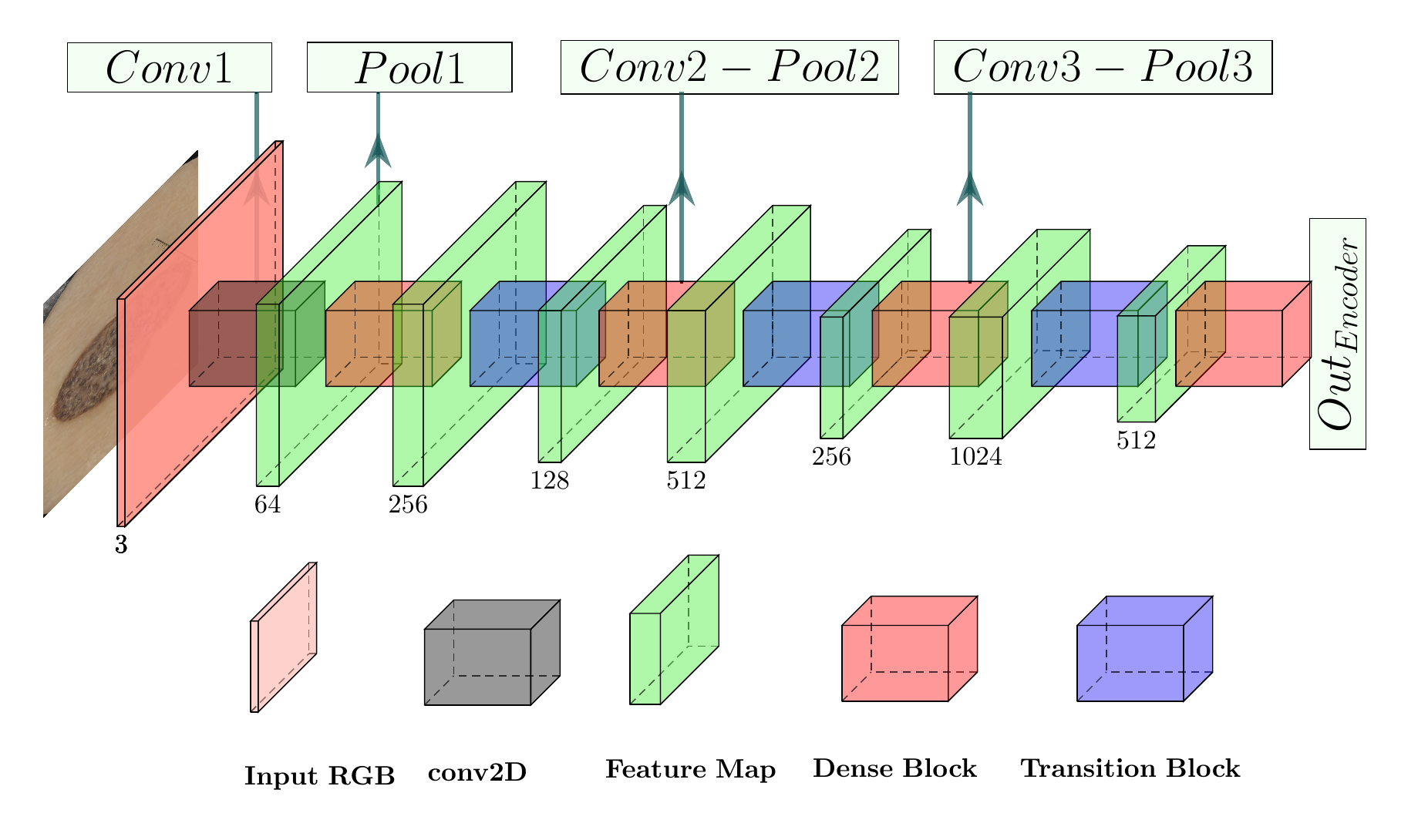}
\caption{The encoder of the DSNet generates spatially invariant feature maps. In a dense block, the feature map resolution remains the same while the number of filters differ whereas in a transition block, the redundant features are truncated.The feature maps named as Conv1, Pool1, Conv2-Pool2, and Conv3-Pool3 will be used for skip connection in deocoder of DSNet.}
\label{fig:encoderb}
\end{figure}

\begin{figure}[t]
\centering
\subfloat{\includegraphics[width=15.5cm,height=8.5cm]{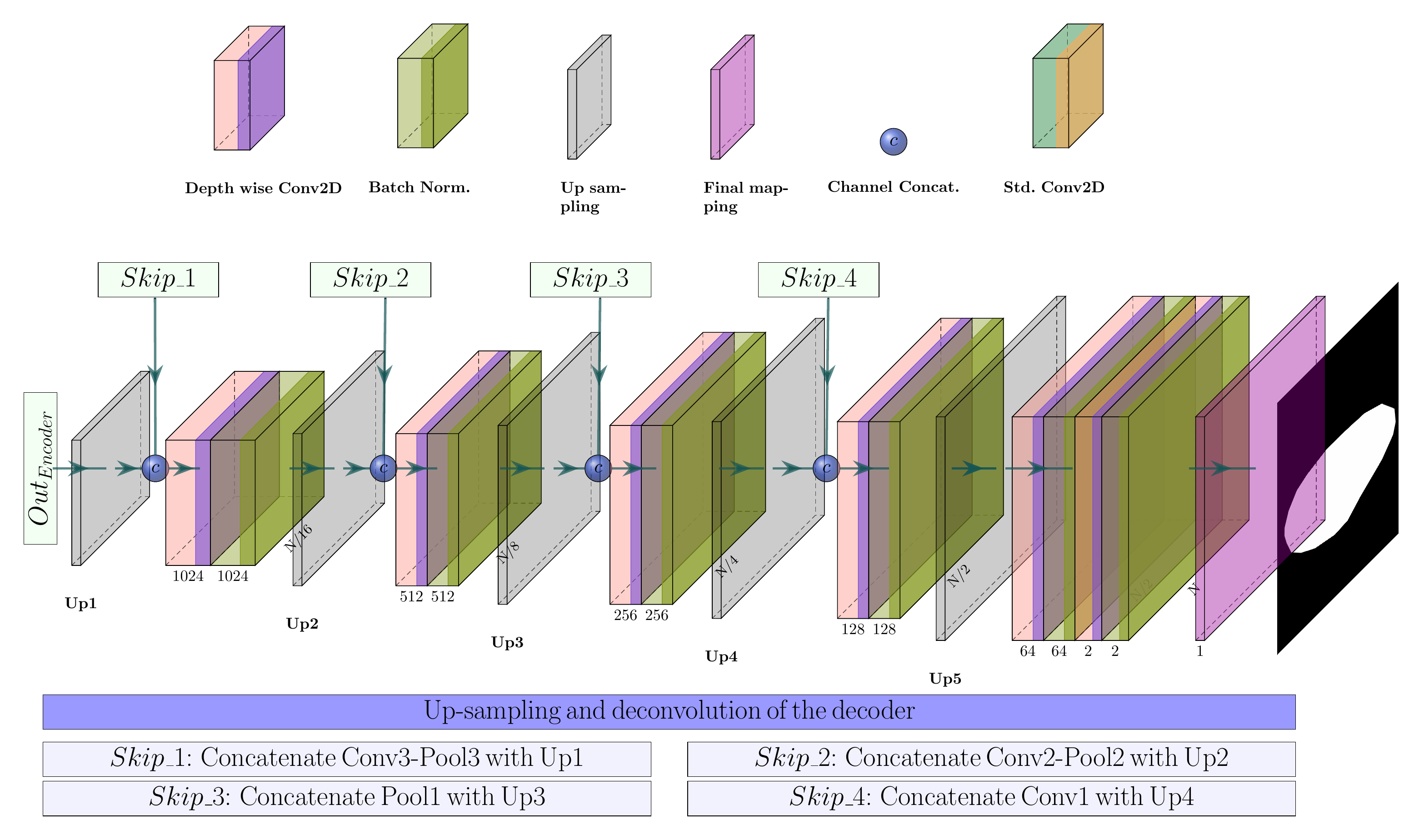}}
\caption{The decoder of the proposed DSNet to reconstruct a high resolution image of the predicted lesion mask from a low resolution image.}
\label{fig:network}
\end{figure}

The encoder of the proposed DSNet consisting of feature layers, dense blocks, and transition blocks. A block representation of the encoder is presented in Fig. \ref{fig:encoderb}. A complete depiction of  all layers and the connectivities (among convolution, batch normalization, activation, deconvolution layers, \textit{etc.}) among them are provided on the GitHub link\footref{note} as it is impractical to do in the article due to the space constraints.
In a dense block, a dense layer receives the outputs of all its previous layers (see Fig. \ref{fig:encodera}) where the output depth of the $L^{th}$ layer $(D_L)$ is \textit{(L-1)} $\times$ \textit{growth rate} $+ D_{L-1}$. The growth rate is the hyperparameter that regulates the amount of information being added to the layer of the network. A transition layer exists between every two adjacent dense layers which is responsible for reducing the computational complexity. In the transition layer the $1 \times 1$ convolutional layer (bottleneck layer)  ensures that the second convolutional layer always has a fixed input depth \citep{denseNet} and the $2 \times 2$ average pooling layer with a stride of $2$ reduces the size of the feature map by sub-sampling. We have also used skip connections having ladder like structures \citep{Ladder} inspired by U-Net to overcome the sub-sampling limitations and \textit{deconvolution overlap}. Each pooled layer of the DSNet is channel-wise concatenated to a deconvoluted feature map having the same dimensions as shown in Fig. \ref{fig:network} where it acts as a compensatory connection for the spatial information lost due to sub-sampling.  




Moreover, the internal co-variate shift due to the change of parameters during the training has been addressed in the proposed DSNet by adding \textit{batch normalization} \citep{Sergey} in both the encoder and decoder. The use of depth-wise separable convolution \citep{Xception} instead of traditional standard convolution in the decoder and the reuse of features that leads to a substantial reduction of the number of parameters in the encoder are the main factors in making the DSNet lightweight. Depth-wise separable convolution is a spatial convolution performed independently over each channel of an input followed by a point-wise convolution, i.e. a $1 \times 1$ convolution that projects each output channel of the depth-wise convolution onto a new channel space. For any convolution layer, suppose the number of filters, depth and kernel size are $N_F$, $M_D$ and $K$ respectively. The the total number of parameters of that layer is $N_F \times M_D \times K^2$ and $M_D \times (N_F+K^2)$ for standard and depth-wise separable convolution respectively. Thus, we are able to reduce the number of parameters by a factor of $(1/N_F + 1/K^2)$ per each convolution layer of DSNet.

\subsection{Datasets and Hardware}
\label{dataset}
The proposed network was trained on the publicly available International Skin Imaging Collaboration (ISIC-2017) training dataset \citep{ISIC2017data} and was tested on both ISIC-2017 test and PH2 \citep{ph2data} datasets. The ISIC-2018 challenge dataset \citep{ISIC2017data} was not used as the organizers haven't provided the groundtruth data for validation and testing set. 
The images in ISIC-2017 dataset are $8$-bit RBG having various resolutions ranging from $540 \times 722$ to $4,499 \times 6,748$ pixels whereas the ones in PH2 dataset are $8$-bit RBG having the same resolution of $768 \times 560$ pixels. The training, validation, and test datasets of ISIC comprise $2,000$, $150$, and $600$ images respectively.
Furthermore, the percentage of melanoma (mel), seborrheic keratosis (sk), and nevus (nev) are ($18.8 \,\%$, $12.7 \, \%$, and $68.5 \, \%$), ($20.0 \, \%$, $28.0 \, \%$, and $52.0 \, \%$), and ($19.5 \, \%$, $15.0 \, \%$, and $65.5 \, \%$) in training, validation, and testing respectively. On the other hand, PH2 dataset has $20.0 \, \%$ and $80.0 \, \%$ of mel and nev respectively. A set of images from the ISIC-2017 test dataset, representing various challenges to overcome is presented in Fig.~\ref{fig:TrainingImages}. The networks were implemented using the python programming language in keras framework with tensorflow backend and the experiments were carried out on a desktop running \textit{Ubuntu} 16.04 LTS operating system with the following hardware configuration: Intel\textsuperscript{\tiny\textregistered} Core\textsuperscript{\tiny{TM}} i7-6850K CPU @ $3.60 GHz \times 12$  processor and GeForce GTX 1080/PCle/SSE2 GPU with 8GB GDDR5 memory.

\subsection{Training and Evaluation Metrics}
In the proposed DSNet, the kernels of the encoder were initialized using the pre-trained weights from ImageNet \citep{imagenet} and the kernels of decoder were initialized using the \say{he normal} distribution \citep{henormal}. 
It draws samples from a truncated normal distribution centered on $0$ and the standard deviation is $\sqrt{2 / K}$ where $K$ is the number of input units of the weight tensor.
All input images were standardized and rescaled to $[0 \, 1]$ range before feeding them into the network. As we noticed that most of the images of the ISIC-2017 dermoscopic dataset have a $3:4$ aspect ratio, all images were resized to $192 \times 256$ pixels.
To increase the robustness of the segmentation, geometric augmentations such as rotation, zooming, shifting, and flipping were performed. The ISODATA (Iterative Self-Organizing Data Analysis Technique) \citep{isodata} method was used to threshold the output probability map from the $sigmoid$ activator in order to retrieve the skin lesion ROI. The region having the maximum connected number of pixels was selected as the skin lesion ROI. The binary or categorical cross-entropy functions are widely used as loss functions in both classification and semantic segmentation. However, using binary or categorical cross-entropy as loss functions may lead to bias effects as the size of a lesions is drastically smaller than the size of the background. Hence, we minimize the sum of binary cross-entropy and IoU \citep{jaccard} as the loss function ($L_{seg}$) and is expressed in Eq. \ref{eq:loss}. 

\begin{equation} \label{eq:loss}
L_{seg}(y,\hat{y})= \frac{1}{N} \sum_{i=1}^{N}[y_i \log \hat{y_i} + (1-y_i)\log (1-\hat{y_i}) ] + 1 - \frac{\sum_{i=1}^{N}y_i\times \hat{y_i}}{\sum_{i=1}^{N}y_i+\sum_{i=1}^{N}\hat{y_i}-\sum_{i=1}^{N}y_i\times \hat{y_i}}
\end{equation}

\noindent where, $y$, $\hat{y}$ and $N$ are the true labels, predicted labels and total numbers of pixel respectively. 
The loss function is optimized using $adadelta$ \citep{adadelta} with \textit{initial learning rate} = $1.0$ and \textit{decay factor} = $0.95$. For better optimization, the initial learning rate was reduced when the specified metric had stopped improving. When learning loss stagnated for $8$ epochs, the learning rate was reduced by $40 \%$. 


The mean Intersection over Union (mIoU), mean Sensitivity (mSn), and mean Specificity (mSp) have been used for the quantitative analysis of segmented masks in different experiments. mIoU is used to quantify the percentage overlap between the true and predicted lesion masks whereas mSn and mSp are used to quantify the type II error (false negative rate) and type I error (false positive rate) respectively. 

\section{Results and Discussion}
\label{Experiments}

In this section, the qualitative and quantitative segmentation results obtained by several extensive experiments are reported.
First, the proposed DSNet is compared against two other implemented networks: namely, FCN8s and U-Net along with a comparison of the loss function ($L_{seg}$) against cross-entropy and intersection over union (IoU).
Then, the proposed DSNet is further evaluated with respect to different classes (mel, sk, and nev) segmentation for both ISIC-2017 and PH2 datasets. 
Finally, the performance of DSNet is compared against the state-of-the-art networks. 



\subsection{Evaluation of Proposed DSNet and Loss function} 




The quantitative results of semantic segmentation on ISIC-2017 test dataset for the three networks (DSNet, FCN8s, and U-Net) employing different loss functions are shown in Table \ref{tab:quancompare}. 
It is observed that irrespective of the segmentation network, the specificity of both the cross entropy and IoU loss functions is higher than the sensitivity which implies that the background skin is segmented more accurately than the lesions. The true positive rate (sensitivity) was improved by using the proposed loss function compared to the IoU loss funtion while the true negative rate (specificity) was improved using the proposed loss function compared to the cross entropy loss function as presented in the Table \ref{tab:quancompare}. The proposed loss function resulted in a larger overlap (mIoU) between the true and predicted masks compared to both cross-entropy and IoU loss functions for all three networks. Furthermore, it is also observed that the proposed DSNet yields the best performances with respect to mIoU by improving the U-Net and FCN8s results 
by $3.6\,\%$ and $6.8\,\%$ respectively when the proposed loss function is employed.
The compactness of DSNet is evidenced by having $3.8$ and $13.8$ times less number of parameters than U-Net and FCN8s respectively. 

\begin{table}[!ht]
    \caption {The quantitative results of semantic segmentation for the three networks (FCN8s, U-Net, and the proposed DSNet) using different loss functions for ISIC-2017 test dataset.
    Metrics were calculated using the ground truth and the predicted labels obtained using the networks.} 
    \label{tab:quancompare} 

    \centering
\begin{tabular}{cccccc}
\hline
\rowcolor[HTML]{C0C0C0} 
\cellcolor[HTML]{C0C0C0} & \cellcolor[HTML]{C0C0C0} & \cellcolor[HTML]{C0C0C0} & \multicolumn{3}{c}{\cellcolor[HTML]{C0C0C0} Metric} \\ \cline{4-6} 
\rowcolor[HTML]{C0C0C0} 
\multirow{-2}{*}{\cellcolor[HTML]{C0C0C0}Network} & \multirow{-2}{*}{\cellcolor[HTML]{C0C0C0}Parameters} & \multirow{-2}{*}{\cellcolor[HTML]{C0C0C0}loss function} & mIoU & mSn & mSp \\ \hline
 &  & \cellcolor[HTML]{DAE8FC}Cross Entropy & \cellcolor[HTML]{DAE8FC}0.688 & \cellcolor[HTML]{DAE8FC}0.926 & \cellcolor[HTML]{DAE8FC}0.893 \\
 &  & \cellcolor[HTML]{FFFFC7}IoU Loss & \cellcolor[HTML]{FFFFC7}0.658 & \cellcolor[HTML]{FFFFC7}0.718 & \cellcolor[HTML]{FFFFC7}0.965 \\
\multirow{-3}{*}{FCN8s} & \multirow{-3}{*}{138 M} & \cellcolor[HTML]{96FFFB}Proposed Loss & \cellcolor[HTML]{96FFFB}0.707 & \cellcolor[HTML]{96FFFB}0.858 & \cellcolor[HTML]{96FFFB}0.938 \\ \hline
 &  & \cellcolor[HTML]{DAE8FC}Cross Entropy & \cellcolor[HTML]{DAE8FC}0.717 & \cellcolor[HTML]{DAE8FC}0.900 & \cellcolor[HTML]{DAE8FC}0.933 \\
 &  & \cellcolor[HTML]{FFFFC7}IoU Loss & \cellcolor[HTML]{FFFFC7}0.693 & \cellcolor[HTML]{FFFFC7}0.719 & \cellcolor[HTML]{FFFFC7}0.986 \\
\multirow{-3}{*}{U-Net} & \multirow{-3}{*}{38 M} & \cellcolor[HTML]{96FFFB}Proposed Loss & \cellcolor[HTML]{96FFFB}0.739 & \cellcolor[HTML]{96FFFB}0.785 & \cellcolor[HTML]{96FFFB}0.982 \\ \hline
 &  & \cellcolor[HTML]{DAE8FC}Cross Entropy & \cellcolor[HTML]{DAE8FC}0.771 & \cellcolor[HTML]{DAE8FC}0.832 & \cellcolor[HTML]{DAE8FC}0.977 \\
 &  & \cellcolor[HTML]{FFFFC7}IoU Loss & \cellcolor[HTML]{FFFFC7}0.743 & \cellcolor[HTML]{FFFFC7}0.782 & \cellcolor[HTML]{FFFFC7}0.984 \\
\multirow{-3}{*}{Proposed Network} & \multirow{-3}{*}{10 M} & \cellcolor[HTML]{96FFFB}Proposed Loss & \cellcolor[HTML]{96FFFB}0.775 & \cellcolor[HTML]{96FFFB}0.875 & \cellcolor[HTML]{96FFFB}0.955 \\ \hline
\end{tabular}
\end{table}

\begin{figure}[!ht]  
\centering 
\scriptsize

{
\begin{tabular}{c c c c}

\textbf{Original} & \textbf{FCN8s} & \textbf{U-Net} & \textbf{Proposed Net} \\ 

\hline
\includegraphics[width=3.7cm,height=2.3cm]{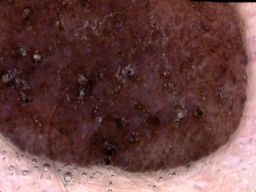}   &
\includegraphics[width=3.7cm,height=2.3cm]{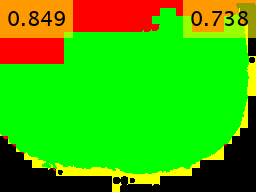}   &
\includegraphics[width=3.7cm,height=2.3cm]{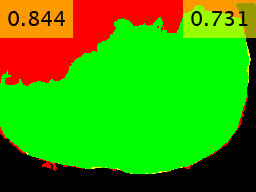}   &
\includegraphics[width=3.7cm,height=2.3cm]{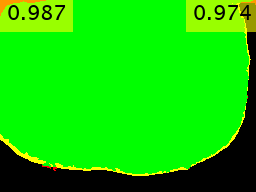}   \\ 

\hline
\includegraphics[width=3.7cm,height=2.3cm]{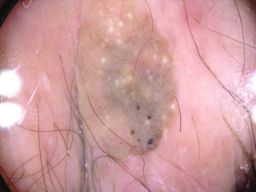}   &
\includegraphics[width=3.7cm,height=2.3cm]{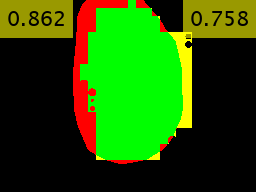}   &
\includegraphics[width=3.7cm,height=2.3cm]{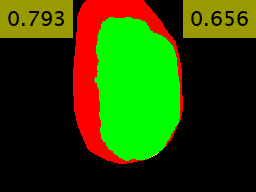}   &
\includegraphics[width=3.7cm,height=2.3cm]{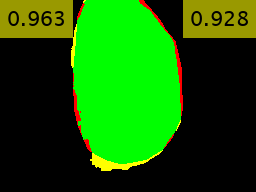}   \\

\hline
\includegraphics[width=3.7cm,height=2.3cm]{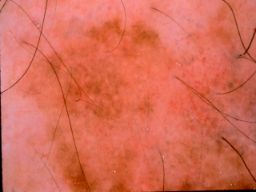}   &
\includegraphics[width=3.7cm,height=2.3cm]{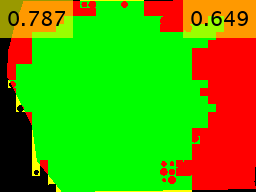}   &
\includegraphics[width=3.7cm,height=2.3cm]{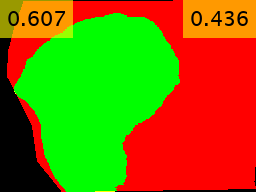}   &
\includegraphics[width=3.7cm,height=2.3cm]{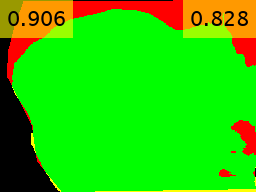}   \\

\hline
\includegraphics[width=3.7cm,height=2.3cm]{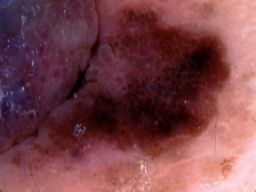}   &
\includegraphics[width=3.7cm,height=2.3cm]{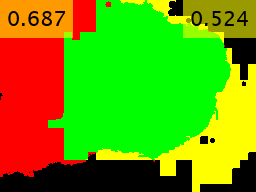}   &
\includegraphics[width=3.7cm,height=2.3cm]{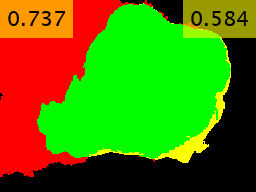}   &
\includegraphics[width=3.7cm,height=2.3cm]{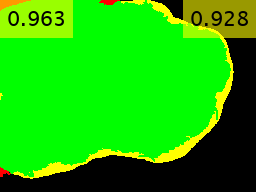}   \\ 

\end{tabular}
}  
\caption[] 
{ Qualitative segmentation results for ISIC-2017 test dataset using FCN8s, U-Net, and DSNet network implementations. Green, Red, and Yellow colors indicate the TP, FN, and FP respectively. The Dice coefficient (top-left) and IoU (top-right) are provided for quantitative evaluation. More segmented results are available on GitHub\footref{note}.}
\label{fig:modelcompare} 
\end{figure}

Fig. \ref{fig:modelcompare} presents a sample of the qualitative results obtained by the three different networks using the proposed loss function.
The checkerboard effect can be observed on the segmented masks of  images obtained using FCN8s due to the non-divisible upscaling factor by the size of the kernels in its decoder when projecting  the lower resolution discriminating features learnt by the encoder onto the pixel space of higher resolution. It can be also noticed that the segmented masks using FCN8s have more False Positives (FP) and coarse boundaries due to \textit{deconvolution overlapping} in the decoder. As evidenced by Fig. \ref{fig:modelcompare}, the masks obtained using the U-Net suffer from under segmentation due to the lack of learnt salient features and lost spatial resolution in the encoder. It can also be clearly seen that there are more False Negatives (FN) with the results obtained using U-Net compared to both FCN8s and DSNet which is highly undesirable in medical diagnosis because skin lesions would be predicted as healthy (background) skin by the CAD system.
On the other hand, the proposed network with the proposed loss function demonstrates better performance in skin lesion segmentation as shown in Fig. \ref{fig:modelcompare}.

\begin{figure}[!ht]
\centering
\includegraphics[width=11cm,keepaspectratio=true]{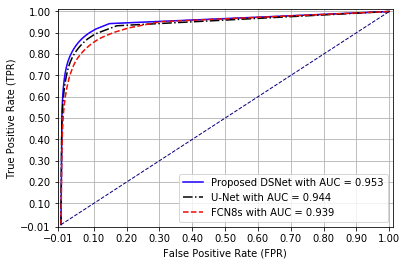}
\caption{ Receiver Operating Characteristic (ROC) of the segmented masks using DSNet, FCN8s, and U-Net. The dashed diagonal line represents the ROC curve of a random predictor as a baseline.}
\label{fig:roc_compare}
\end{figure}

The proposed approach obtains a lower number of false positives and false negatives compared to the other two implemented networks. 
Fig. \ref{fig:roc_compare} shows receiver operating characteristic (ROC) curve of the different networks. 
The proposed DSNet obtains an area under ROC curve (AUC) of $0.953$ which indicates that for any given random pixel, the probability of accurate classification is as high as $95.3 \,\%$. 
Also from Fig. \ref{fig:roc_compare} and given a $10 \, \%$ false positive rate, the true positive rates of the FCN8s, U-Net, and proposed DSNet are $86.0 \, \%$, $90.0 \, \%$, and $92.0 \, \%$ respectively. 
The fact that the proposed DSNet not only learns highly abstract salient features of skin lesions using a lower number of parameters (10 M compared to 38 M and 138 M of FCN8s and U-Net respectively) but also leads to improved semantic segmentation performance indicate that it can be potentially used as a CAD tool for lesion segmentation. 
In some cases as shown in \ref{appendix1}, DSNet also fails to segment certain skin lesions. However, in those wrong segmentations, the false positive rate is generally more than the false negative rate which indicates a better suitability of DSNet for medical applications.


\subsection{DSNet Performance on Different Classes }

\begin{figure}[!ht]  
\centering  
  {\scriptsize 
  \setlength\tabcolsep{4pt} 
\begin{tabular}{cc||cc}
\multicolumn{2}{c||}{\textbf{ISIC-2017 test dataset}} & \multicolumn{2}{c}{\textbf{PH2 dataset}}  \\ \cline{1-4} 
\textbf{Original} &  \textbf{Overlaid} & \textbf{Original}  & \textbf{Overlaid}\\ 
\hline
\includegraphics[width=3.45cm,height=2.4cm]{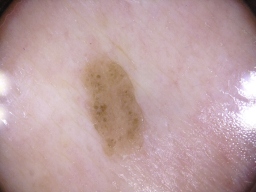}   &
\includegraphics[width=3.45cm,height=2.4cm]{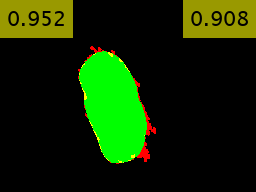}   &
\includegraphics[width=3.45cm,height=2.4cm]{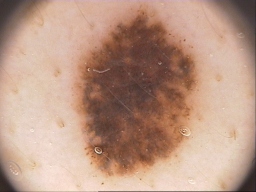}   &
\includegraphics[width=3.45cm,height=2.4cm]{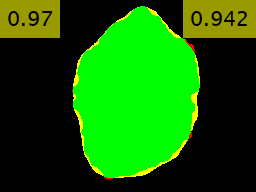}   \\ 
\hline
\includegraphics[width=3.45cm,height=2.4cm]{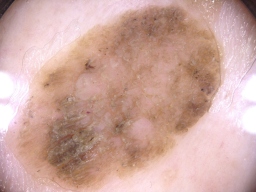}   &
\includegraphics[width=3.45cm,height=2.4cm]{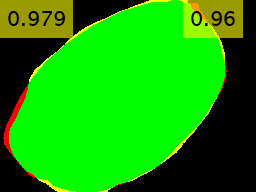}   &
\includegraphics[width=3.45cm,height=2.4cm]{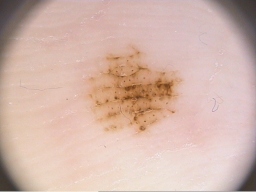}   &
\includegraphics[width=3.45cm,height=2.4cm]{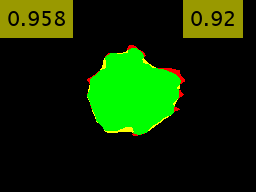}   \\ 
\hline
\includegraphics[width=3.45cm,height=2.4cm]{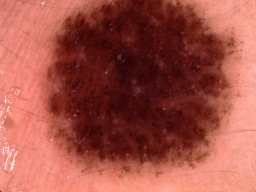}   &
\includegraphics[width=3.45cm,height=2.4cm]{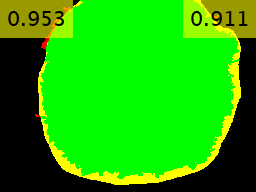}   &
\includegraphics[width=3.45cm,height=2.4cm]{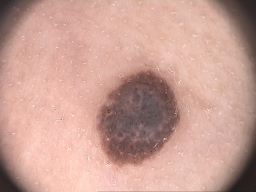}   &
\includegraphics[width=3.45cm,height=2.4cm]{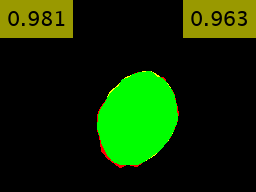}   \\ 
\hline
\includegraphics[width=3.45cm,height=2.4cm]{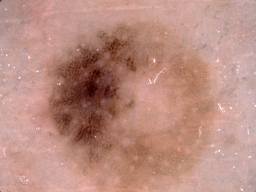}   &
\includegraphics[width=3.45cm,height=2.4cm]{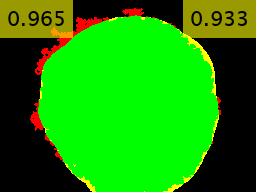}   &
\includegraphics[width=3.45cm,height=2.4cm]{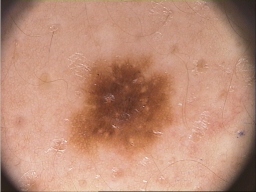}   &
\includegraphics[width=3.45cm,height=2.4cm]{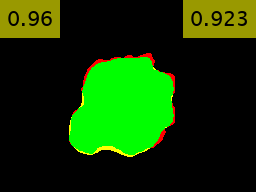}   \\ 
\hline
\end{tabular}
}  
\caption[]{Qualitative segmentation results for ISIC-2017 test and PH2 datasets using the proposed DSNet network. Green, Red, and Yellow colors indicate the TP, FN, and FP respectively.
The Dice co-efficient (top-left) and IoU (top-right) are provided for quantitative evaluation.
More segmented results are available on GitHub\footref{note}.}
\label{fig:quali2}
\end{figure}

The proposed DSNet was further evaluated on different classes of skin lesion for both ISIC-2017 test and PH2 datasets. The quantitative and qualitative results for mel, sk, and nev lesion class segmentation are presented in Table \ref{tab:quan2} and Fig. \ref{fig:quali2} respectively. 
\begin{table}[!ht]
\caption {The quantitative results for different skin lesion class segmentation for both ISIC-2017 test and PH2 datasets using the proposed DSNet. Metrics were calculated using the ground truth and the predicted labels obtained using the networks.}
 \centering   
\begin{tabular}{c
>{\columncolor[HTML]{FFFFC7}}c 
>{\columncolor[HTML]{FFFFC7}}c 
>{\columncolor[HTML]{FFFFC7}}c 
>{\columncolor[HTML]{FFFFC7}}c 
>{\columncolor[HTML]{CBCEFB}}c 
>{\columncolor[HTML]{CBCEFB}}c 
>{\columncolor[HTML]{CBCEFB}}c }
\hline
\cellcolor[HTML]{C0C0C0} & \multicolumn{4}{c}{\cellcolor[HTML]{C0C0C0}ISIC-2017 test dataset} & \multicolumn{3}{c}{\cellcolor[HTML]{C0C0C0}PH2 dataset} \\ \cline{2-8} 
\multirow{-2}{*}{\cellcolor[HTML]{C0C0C0}Metric} & \cellcolor[HTML]{C0C0C0}nev & \cellcolor[HTML]{C0C0C0}mel & \cellcolor[HTML]{C0C0C0}sk & \cellcolor[HTML]{C0C0C0}overall & \cellcolor[HTML]{C0C0C0}nev & \cellcolor[HTML]{C0C0C0}mel & \cellcolor[HTML]{C0C0C0}overall \\ \hline
    mIoU     & 0.808     & 0.730     & 0.684     & 0.775   &  0.891     & 0.835      & 0.870    \\ \hline
    mSn      & 0.907     & 0.836     & 0.832     & 0.875   &  0.945     &  0.929     & 0.929     \\ \hline
    mSp      & 0.956     & 0.939     & 0.953     & 0.955   &  0.976     &  0.849      & 0.969        \\ \hline
\end{tabular}
\label{tab:quan2}
\end{table}
\begin{figure}[!ht]
\centering
\subfloat[]{\includegraphics[width=8cm,height=5cm]{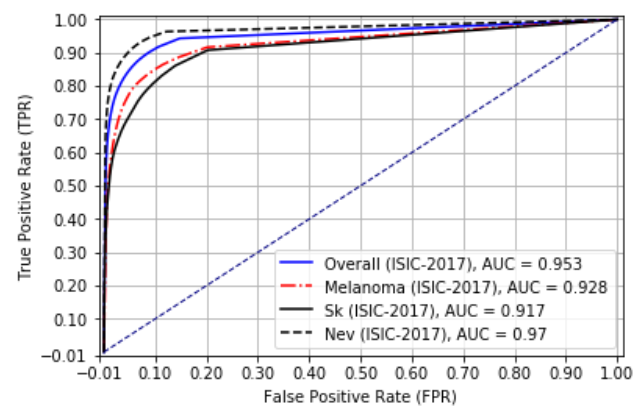}}
\subfloat[]{\includegraphics[width=8cm,height=5cm]{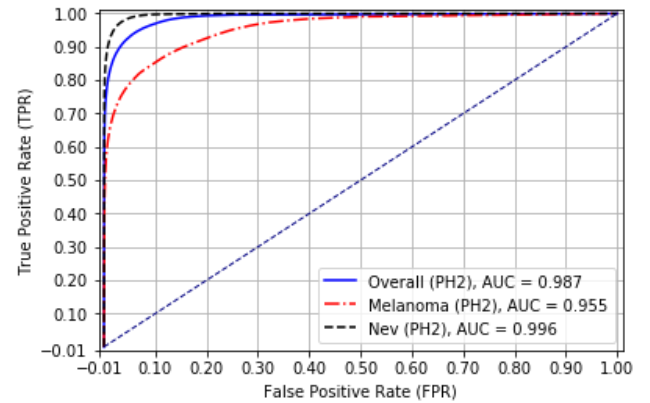}}
\caption{Class wise ROC curves for lesion segmentation using the proposed DSNet. a) ROC curves for mel, sk, and nev classes for ISIC-2017 test dataset and b) ROC curves for mel and nev classes for PH2 dataset.}
\label{fig:roc_compare2}
\end{figure}
Even though the training images were imbalanced for different lesion classes, the high quality segmentation results obtained for all classes (see Table \ref{tab:quan2}) a better performance of our proposed network for  semantic lesion segmentation. The qualitative results in Fig. \ref{fig:quali2} show the accuracy of the lesion segmentation using DSNet for both ISIC-2017 test and PH2 datasets although PH2 dataset was neither used for training nor validation of DSNet. The mIoU ($0.87$) on PH2 demonstrates the robustness of the proposed DSNet since the most abstract features were learnt from the ISIC-2017.  
From Table \ref{tab:quan2}, it can be also observed that for both ISIC and PH2 datasets, the nev class is easier to segment than other two skin lesion classes due to the larger existing percentage of nev cases in the training dataset (there are $3.7$ and $5.4$ times more nev images than mel and sk classes respectively).
The ROC curves obtained from the segmented lesion masks using the proposed DSNet for different lesion classes of both datasets are presented in Fig. \ref{fig:roc_compare2}.
In can be observed that for $10 \, \%$ false positive rate,  the true positive rates of sk, mel, and nev classes respectively are $82.0 \, \%$, $85.0 \, \%$, and $95.0 \, \%$ for ISIC-2017 test dataset while the true positive rates of mel and nev classes are $85.0 \, \%$ and $99.0 \, \%$ respectively for PH2 dataset. 
These segmentation mask results on both ISIC and PH2 datasets vouch for the suitability of DSNet for different skin lesion class segmentation and its success as a CAD tool for melanoma detection in CAD systems.

\subsection{Advantages of DSNet over Recent Networks}
In this section, the performance of DSNet is compared and contrasted to several recent networks used for skin lesion segmentation using the same datasets for training, validation, and testing. The AUC of the segmented masks for proposed DSNet and FrCN \citep{Mohammed} are provided in Table \ref{tab:AUC} which were trained and tested on the same datasets. DSNet outperforms FrCN by substantial margins in all lesion class segmentation for both ISIC-2017 test and PH2 datasets. Consequently, these results indicate the ability of DSNet to learn more salient and discriminant features of skin lesions comparative to FrCN.

\begin{table}[!ht]
\caption {The AUC of the DSNet and the recent FrCN by \citet{Mohammed} for skin lesion class segmentation for ISIC-2017 test and PH2 datasets.
Both networks were trained and tested on the same datasets.}
\label{tab:AUC} 
\centering
\begin{tabular}{cccccccc}
\hline
\rowcolor[HTML]{C0C0C0} 
\cellcolor[HTML]{C0C0C0} & \multicolumn{4}{c}{\cellcolor[HTML]{C0C0C0}ISIC-2017 test dataset} & \multicolumn{3}{c}{\cellcolor[HTML]{C0C0C0}PH2 dataset} \\ \cline{2-8} 
\rowcolor[HTML]{C0C0C0} 
\multirow{-2}{*}{\cellcolor[HTML]{C0C0C0}Network} & mel & sk & nev & overall & mel & nev & overall \\ \hline
FrCN & 0.875 & 0.882 & 0.932 & 0.910 & 0.941 & 0.950 & 0.946 \\ \hline
\textbf{DSNet (Proposed)} & \textbf{0.928} & \textbf{0.917} & \textbf{0.970} & \textbf{0.953} & \textbf{0.955} & \textbf{0.996} & \textbf{0.987} \\ \hline
\end{tabular}
\end{table}

Skin lesion segmentation performance of the proposed DSNet and a number of other state-of-the-art networks for ISIC-2017 test and PH2 datasets are presented in Table \ref{tab:segmethodscompare}. DSNet produces the best segmentation for three out of the six cases while performing similarly with the winning networks on the other three cases. DSNet produces the best results with respect to the true positive rate (lesion detected as lesion) beating the state-of-the-art network \citep{Shen} by a $2.0 \, \%$ margin. With respect to the true negative rate, DSNet is behind the state-of-the-art network MResNet-Seg \citep{r24} by $3.0 \, \%$. However, comparing the true positive rate of DSNet (0.875) and MresNet-Seg (0.802), we observe that DSNet outperforms the latter by a $7.3 \, \%$ margin providing ample evidence for the higher lesion region segmentation ability of the former. With respect to mIoU, DSNet produces the second best results behind SLSDeep \citep{r18} by a small margin of $0.7\,\%$ indicating that the latter network is able to learn the background more accurately than the former. However, DSNet outperforms SLSDeep with respect to the true positive rate by a margin of $5.9\,\%$ demonstrating the higher ability of the former to correctly segment skin lesions. Additionally, the proposed DSNet is a lighter network compared to SLSDeep as the former has less number parameters than the latter. Moreover, the proposed network demonstrates excellent performance on PH2 dataset than the state-of-the-art networks as presented in Table \ref{tab:segmethodscompare}. Since the proposed network also performs better on PH2 dataset, it can be claimed that DSNet is more generic as it possesses the capacity to learn more abstract features by efficient reuse of the features inside the dense blocks in the encoder. 

\begin{table}[!ht]
\caption {Semantic segmentation performance metrics for the proposed DSNet and other state-of-the-art networks on both ISIC-2017 test and PH2 datasets.}
\centering
\label{tab:segmethodscompare} 
  \begin{tabular}{ c || c c c || c  c c c}
        \hline
        \rowcolor[HTML]{C0C0C0} 
        \cellcolor[HTML]{C0C0C0}{\color[HTML]{000000} } & \multicolumn{3}{c||}{\cellcolor[HTML]{C0C0C0}ISIC-2017 test dataset} & \multicolumn{3}{c}{\cellcolor[HTML]{C0C0C0}PH2 dataset} \\ \cline{2-7} 
        \rowcolor[HTML]{C0C0C0} 
        \multirow{-2}{*}{\cellcolor[HTML]{C0C0C0}{\color[HTML]{000000} Network}} & mIoU & mSn & mSp & mIoU & mSn & mSp \\ \hline
        SegNet\textsuperscript{*} \citep{Mohammed} & 0.696  & 0.801 & 0.954  & 0.808 & 0.865 & 0.966  \\ \hline
        II-FCN\citep{Wen} &  0.699 & 0.841  &  0.984  &-  &-  &-  \\ \hline
        LIN \citep{Shen} & 0.753 & 0.855 &  0.974 &-  &-  &-  \\ \hline
        MResNet-Seg \citep{r24} & 0.760 & 0.802 & \textbf{0.985} &-  &-  & - \\ \hline
        FrCN \citep{Mohammed} & 0.771  & 0.854 & 0.967 & 0.848 & \textbf{0.937} & 0.957 \\ \hline
        SLSDeep \citep{r18} & \textbf{0.782}  & 0.816 & 0.983 & - & - & - \\ \hline
        \textbf{Proposed DSNet}    &  0.775     & \textbf{0.875}    & 0.955  &  \textbf{0.870}     & 0.929    & \textbf{0.969} \\ \hline
  \end{tabular}
   {\raggedright \textsuperscript{*}\footnotesize{Originally proposed by \citet{Vijay} and was implemented by \citet{Mohammed} for skin lesion segmentation.\par}}
\end{table}

Additionally, the average test time for a single dermoscopy image was $0.595\,sec$ (357 seconds for 600 images) for the proposed DSNet (10M parameters) compared to the $9.7\,sec$ for FrCN (184M parameters) by \citet{Mohammed}.
The proposed DSNet appears better suited for eventual CAD systems due to its extremely fast processing time compared to the current state-of-the art (roughly 16 times faster than recent FrCN).

The following presents a concise discussion of the possible bottlenecks that may lead the recent networks shown in Table \ref{tab:segmethodscompare} to perform poorly comparatively to the proposed DSNet. Pooling layers are required to reduce the in-plane dimensionality in order to introduce translation invariance, to decrease the number of subsequent learnable parameters, and to increase the capability of learning more abstract features \citep{Rikiya}. However, FrCN does not have pooling layers to capture the full resolution feature maps. 
Consequently, this may explain the lower AUC values obtained.
In SegNet for skin lesion segmentation implemented by \citet{Mohammed}, the authors store the indices at each max-pooling layer in the encoder which are later used to upsample the corresponding feature map in the decoder in order to preserve the high-frequency information. 
Nevertheless, the authors do not take the neighbouring information into account during upsampling. In II-FCN, \citet{Wen} uses inception blocks which are very deep networks and are prone to overfitting. LIN (Lesion Indexing Network) \citep{Shen} is based on fully convolutional residual networks (FCRN-88) and is the deepest network among the networks presented in the Table \ref{tab:segmethodscompare}. The extensive depth of LIN may also lead the CNN model to be overfitted due to the limited numbers of images comparative to its depth. \citet{r24} uses ResNets architecture in MResNet-Seg which often has  gradient fading problems that lead the CNN model to be overfitted. 


\section{Conclusion}
\label{Conclusions}
In this article, a new robust and automatic skin lesion segmentation network called DSNet has been proposed and implemented. Additionally, a new loss function comprising a binary cross-entropy term and intersection over union term has also been proposed. The potential of the proposed network has been evidenced through multiple extensive experiments.
The hurdle of having a limited number of labeled skin lesion images to train the DSNet was overcome by transferring the knowledge from another pre-trained model and by augmenting the images. The proposed new loss function with the addition of the IoU to the more traditional cross-entropy has proven to be better suited for semantic segmentation by achieving higher true positive rates in the experiments carried out. The proposed loss function may play an important role in semantic segmentation where ROI selection is crucial. 
The proposed DSNet possesses the highly desirable trait of having the least number of parameters among all compared networks, while outperforming the existing baseline segmentation networks with respect to several metrics. Thus, DSNet is the least memory intensive among all compared networks. 
Further tuning of the hyper-parameters and the introduction of more color space augmentations may yield better segmentation performances.
In the future, the segmented ROIs can be used to extract features for different lesion type classification (mel, sk, and nev). 
The proposed DSNet and the loss function will be applied to other medical contexts to verify its generality and versatility.


\bibliographystyle{model2-names}

\bibliography{sample}

\begin{thebibliography}{47}
\expandafter\ifx\csname natexlab\endcsname\relax\def\natexlab#1{#1}\fi
\expandafter\ifx\csname url\endcsname\relax
  \def\url#1{\texttt{#1}}\fi
\expandafter\ifx\csname urlprefix\endcsname\relax\def\urlprefix{URL }\fi
\providecommand{\eprint}[2][]{\url{#2}}
\providecommand{\bibinfo}[2]{#2}
\ifx\xfnm\relax \def\xfnm[#1]{\unskip,\space#1}\fi
\bibitem[{Abbas et~al.(2011)Abbas, Fondon and Rashid}]{Abbas}
\bibinfo{author}{Abbas, Q.}, \bibinfo{author}{Fondon, I.},
  \bibinfo{author}{Rashid, M.}, \bibinfo{year}{2011}.
\newblock \bibinfo{title}{Unsupervised skin lesions border detection via
  two-dimensional image analysis}.
\newblock \bibinfo{journal}{Computer methods and programs in biomedicine}
  \bibinfo{volume}{104}, \bibinfo{pages}{1--15}.
\bibitem[{Ajmal et~al.(2018)Ajmal, Rehman, Farooq, Ain, Riaz and Hassan}]{r12}
\bibinfo{author}{Ajmal, H.}, \bibinfo{author}{Rehman, S.},
  \bibinfo{author}{Farooq, U.}, \bibinfo{author}{Ain, Q.U.},
  \bibinfo{author}{Riaz, F.}, \bibinfo{author}{Hassan, A.},
  \bibinfo{year}{2018}.
\newblock \bibinfo{title}{Convolutional neural network based image
  segmentation: a review}, in: \bibinfo{booktitle}{Pattern Recognition and
  Tracking XXIX}, \bibinfo{organization}{International Society for Optics and
  Photonics}. p. \bibinfo{pages}{106490N}.
\bibitem[{Al-masni et~al.(2018)Al-masni, Al-antari, Choi, Han and
  Kim}]{Mohammed}
\bibinfo{author}{Al-masni, M.A.}, \bibinfo{author}{Al-antari, M.A.},
  \bibinfo{author}{Choi, M.}, \bibinfo{author}{Han, S.}, \bibinfo{author}{Kim,
  T.}, \bibinfo{year}{2018}.
\newblock \bibinfo{title}{Skin lesion segmentation in dermoscopy images via
  deep full resolution convolutional networks}.
\newblock \bibinfo{journal}{Computer Methods and Programs in Biomedicine}
  \bibinfo{volume}{162}, \bibinfo{pages}{221--231}.
\bibitem[{{American Cancer Society}(2019)}]{r2}
\bibinfo{author}{{American Cancer Society}}, \bibinfo{year}{2019}.
\newblock \bibinfo{title}{Key statistics for melanoma skin cancer}.
\newblock
  \bibinfo{howpublished}{\url{https://www.cancer.org/cancer/melanoma-skin-cancer/about/key-statistics.html}}.
\newblock \bibinfo{note}{[Online; accessed 10-May-2019]}.
\bibitem[{{Badrinarayanan} et~al.(2017){Badrinarayanan}, {Kendall} and
  {Cipolla}}]{Vijay}
\bibinfo{author}{{Badrinarayanan}, V.}, \bibinfo{author}{{Kendall}, A.},
  \bibinfo{author}{{Cipolla}, R.}, \bibinfo{year}{2017}.
\newblock \bibinfo{title}{{SegNet:} a deep convolutional encoder-decoder
  architecture for image segmentation}.
\newblock \bibinfo{journal}{IEEE Transactions on Pattern Analysis and Machine
  Intelligence} \bibinfo{volume}{39}, \bibinfo{pages}{2481--2495}.
\bibitem[{Bi et~al.(2017a)Bi, Kim, Ahn and Feng}]{r24}
\bibinfo{author}{Bi, L.}, \bibinfo{author}{Kim, J.}, \bibinfo{author}{Ahn, E.},
  \bibinfo{author}{Feng, D.}, \bibinfo{year}{2017}a.
\newblock \bibinfo{title}{Automatic skin lesion analysis using large-scale
  dermoscopy images and deep residual networks}.
\newblock \bibinfo{journal}{arXiv:1703.04197, 2017} .
\bibitem[{Bi et~al.(2017b)Bi, Kim, Ahn, Kumar, Fulham and Feng}]{r16}
\bibinfo{author}{Bi, L.}, \bibinfo{author}{Kim, J.}, \bibinfo{author}{Ahn, E.},
  \bibinfo{author}{Kumar, A.}, \bibinfo{author}{Fulham, M.},
  \bibinfo{author}{Feng, D.}, \bibinfo{year}{2017}b.
\newblock \bibinfo{title}{Dermoscopic image segmentation via multistage fully
  convolutional networks}.
\newblock \bibinfo{journal}{IEEE Transactions on Biomedical Engineering}
  \bibinfo{volume}{64}, \bibinfo{pages}{2065--2074}.
\bibitem[{Celebi et~al.(2008)Celebi, Kingravi, Iyatomi, Aslandogan, Stoecker,
  Moss, Malters, Grichnik, Marghoob and Rabinovitz}]{r7}
\bibinfo{author}{Celebi, E.M.}, \bibinfo{author}{Kingravi, H.A.},
  \bibinfo{author}{Iyatomi, H.}, \bibinfo{author}{Aslandogan, Y.A.},
  \bibinfo{author}{Stoecker, W.V.}, \bibinfo{author}{Moss, R.H.},
  \bibinfo{author}{Malters, J.M.}, \bibinfo{author}{Grichnik, J.M.},
  \bibinfo{author}{Marghoob, A.A.}, \bibinfo{author}{Rabinovitz, H.S.},
  \bibinfo{year}{2008}.
\newblock \bibinfo{title}{Border detection in dermoscopy images using
  statistical region merging}.
\newblock \bibinfo{journal}{Skin Research and Technology} \bibinfo{volume}{14},
  \bibinfo{pages}{347--353}.
\bibitem[{Celebi et~al.(2009)Celebi, Iyatomi, Schaefer and Stoecker}]{r4}
\bibinfo{author}{Celebi, M.E.}, \bibinfo{author}{Iyatomi, H.},
  \bibinfo{author}{Schaefer, G.}, \bibinfo{author}{Stoecker, W.V.},
  \bibinfo{year}{2009}.
\newblock \bibinfo{title}{Lesion border detection in dermoscopy images}.
\newblock \bibinfo{journal}{Computerized medical imaging and graphics}
  \bibinfo{volume}{33}, \bibinfo{pages}{148--153}.
\bibitem[{Chollet(2017)}]{Xception}
\bibinfo{author}{Chollet, F.}, \bibinfo{year}{2017}.
\newblock \bibinfo{title}{Xception: Deep learning with depthwise separable
  convolutions}.
\newblock \bibinfo{journal}{IEEE Conference on Computer Vision and Pattern
  Recognition (CVPR)} , \bibinfo{pages}{1800--1807}.
\bibitem[{Codella et~al.(2018)Codella, Gutman, Celebi, Helba, Marchetti, Dusza,
  Kalloo, Liopyris, Mishra and Kittler}]{ISIC2017data}
\bibinfo{author}{Codella, N.F.}, \bibinfo{author}{Gutman, D.},
  \bibinfo{author}{Celebi, M.E.}, \bibinfo{author}{Helba, B.},
  \bibinfo{author}{Marchetti, M.A.}, \bibinfo{author}{Dusza, S.W.},
  \bibinfo{author}{Kalloo, A.}, \bibinfo{author}{Liopyris, K.},
  \bibinfo{author}{Mishra, N.}, \bibinfo{author}{Kittler, H.},
  \bibinfo{year}{2018}.
\newblock \bibinfo{title}{Skin lesion analysis toward melanoma detection: A
  challenge at the 2017 international symposium on biomedical imaging (isbi),
  hosted by the international skin imaging collaboration (isic)}, in:
  \bibinfo{booktitle}{2018 IEEE 15th International Symposium on Biomedical
  Imaging (ISBI 2018)}, \bibinfo{organization}{IEEE}. pp.
  \bibinfo{pages}{168--172}.
\bibitem[{Deng et~al.(2009)Deng, Dong, Socher, Li, Li and Fei-Fei}]{imagenet}
\bibinfo{author}{Deng, J.}, \bibinfo{author}{Dong, W.},
  \bibinfo{author}{Socher, R.}, \bibinfo{author}{Li, L.}, \bibinfo{author}{Li,
  K.}, \bibinfo{author}{Fei-Fei, L.}, \bibinfo{year}{2009}.
\newblock \bibinfo{title}{{ImageNet}: A large-scale hierarchical image
  database}, \bibinfo{publisher}{IEEE Conference on Computer Vision and Pattern
  Recognition}. pp. \bibinfo{pages}{248--255}.
\bibitem[{Doi(2007)}]{cad}
\bibinfo{author}{Doi, K.}, \bibinfo{year}{2007}.
\newblock \bibinfo{title}{Computer-aided diagnosis in medical imaging:
  Historical review, current status and future potential}.
\newblock \bibinfo{journal}{Comput Med Imaging Graph} \bibinfo{volume}{31},
  \bibinfo{pages}{198--211}.
\bibitem[{El-Zaart(2010)}]{isodata}
\bibinfo{author}{El-Zaart, A.}, \bibinfo{year}{2010}.
\newblock \bibinfo{title}{Images thresholding using isodata technique with
  gamma distribution}.
\newblock \bibinfo{journal}{Pattern Recognition and Image Analysis}
  \bibinfo{volume}{20}, \bibinfo{pages}{29--41}.
\bibitem[{Esteva et~al.(2017)Esteva, Kuprel, Novoato, Ko, Swetter, Blau and
  Thrun}]{esteva2017dermatologist}
\bibinfo{author}{Esteva, A.}, \bibinfo{author}{Kuprel, B.},
  \bibinfo{author}{Novoato, R.A.}, \bibinfo{author}{Ko, J.},
  \bibinfo{author}{Swetter, S.M.}, \bibinfo{author}{Blau, H.M.},
  \bibinfo{author}{Thrun, S.}, \bibinfo{year}{2017}.
\newblock \bibinfo{title}{Dermatologist-level classification of skin cancer
  with deep neural networks}.
\newblock \bibinfo{journal}{Nature} \bibinfo{volume}{542},
  \bibinfo{pages}{115--118}.
\bibitem[{Fan et~al.(2017)Fan, Xie, Li, Jiang and Liu}]{r5}
\bibinfo{author}{Fan, H.}, \bibinfo{author}{Xie, F.}, \bibinfo{author}{Li, Y.},
  \bibinfo{author}{Jiang, Z.}, \bibinfo{author}{Liu, J.}, \bibinfo{year}{2017}.
\newblock \bibinfo{title}{Automatic segmentation of dermoscopy images using
  saliency combined with otsu threshold}.
\newblock \bibinfo{journal}{Computers in biology and medicine}
  \bibinfo{volume}{85}, \bibinfo{pages}{75--85}.
\bibitem[{Garcia-Garcia et~al.(2018)Garcia-Garcia, Orts-Escolano, Oprea,
  Villena-Martinez, Martinez-Gonzalez and Garcia-Rodriguez}]{Garcia}
\bibinfo{author}{Garcia-Garcia, A.}, \bibinfo{author}{Orts-Escolano, S.},
  \bibinfo{author}{Oprea, S.}, \bibinfo{author}{Villena-Martinez, V.},
  \bibinfo{author}{Martinez-Gonzalez, P.}, \bibinfo{author}{Garcia-Rodriguez,
  J.}, \bibinfo{year}{2018}.
\newblock \bibinfo{title}{A survey on deep learning techniques for image and
  video semantic segmentation}.
\newblock \bibinfo{journal}{Applied Soft Computing} \bibinfo{volume}{70},
  \bibinfo{pages}{41--65}.
\bibitem[{Garnavi et~al.(2010)Garnavi, Aldeen, Celebi, Bhuiyan, Dolianitis and
  Varigos}]{r6}
\bibinfo{author}{Garnavi, R.}, \bibinfo{author}{Aldeen, M.},
  \bibinfo{author}{Celebi, M.E.}, \bibinfo{author}{Bhuiyan, A.},
  \bibinfo{author}{Dolianitis, C.}, \bibinfo{author}{Varigos, G.},
  \bibinfo{year}{2010}.
\newblock \bibinfo{title}{Automatic segmentation of dermoscopy images using
  histogram thresholding on optimal color channels}.
\newblock \bibinfo{journal}{International Journal of Medicine and Medical
  Sciences} \bibinfo{volume}{1}, \bibinfo{pages}{126--134}.
\bibitem[{Ge et~al.(2017)Ge, Demyanov, Chakravorty, Bowling and Garnavi}]{r1}
\bibinfo{author}{Ge, Z.}, \bibinfo{author}{Demyanov, S.},
  \bibinfo{author}{Chakravorty, R.}, \bibinfo{author}{Bowling, A.},
  \bibinfo{author}{Garnavi, R.}, \bibinfo{year}{2017}.
\newblock \bibinfo{title}{Skin disease recognition using deep saliency features
  and multimodal learning of dermoscopy and clinical images}, in:
  \bibinfo{booktitle}{International Conference on Medical Image Computing and
  Computer-Assisted Intervention}, pp. \bibinfo{pages}{250--258}.
\bibitem[{Guo et~al.(2018)Guo, Liu, Georgiou and Lew}]{r13}
\bibinfo{author}{Guo, Y.}, \bibinfo{author}{Liu, Y.},
  \bibinfo{author}{Georgiou, T.}, \bibinfo{author}{Lew, M.S.},
  \bibinfo{year}{2018}.
\newblock \bibinfo{title}{A review of semantic segmentation using deep neural
  networks}.
\newblock \bibinfo{journal}{International journal of multimedia information
  retrieval} \bibinfo{volume}{7}, \bibinfo{pages}{87--93}.
\bibitem[{Haenssle et~al.(2018)Haenssle, Fink, Schneiderbauer, Toberer, Buhl,
  Blum, Kalloo, Hassen, Thomas and Enk}]{dermatologist}
\bibinfo{author}{Haenssle, H.A.}, \bibinfo{author}{Fink, C.},
  \bibinfo{author}{Schneiderbauer, R.}, \bibinfo{author}{Toberer, F.},
  \bibinfo{author}{Buhl, T.}, \bibinfo{author}{Blum, A.},
  \bibinfo{author}{Kalloo, A.}, \bibinfo{author}{Hassen, A.},
  \bibinfo{author}{Thomas, L.}, \bibinfo{author}{Enk, A.},
  \bibinfo{year}{2018}.
\newblock \bibinfo{title}{Man against machine: diagnostic performance of a deep
  learning convolutional neural network for dermoscopic melanoma recognition in
  comparison to 58 dermatologists}.
\newblock \bibinfo{journal}{Annals of Oncology} \bibinfo{volume}{29},
  \bibinfo{pages}{1836--1842}.
\bibitem[{He et~al.(2015)He, Zhang, Ren and Sun}]{henormal}
\bibinfo{author}{He, K.}, \bibinfo{author}{Zhang, X.}, \bibinfo{author}{Ren,
  S.}, \bibinfo{author}{Sun, J.}, \bibinfo{year}{2015}.
\newblock \bibinfo{title}{Delving deep into rectifiers: Surpassing human-level
  performance on imagenet classification}.
\newblock \bibinfo{journal}{arXiv:1502.01852} .
\bibitem[{{Huang} et~al.(2017){Huang}, {Liu}, v.~d. {Maaten} and
  {Weinberger}}]{denseNet}
\bibinfo{author}{{Huang}, G.}, \bibinfo{author}{{Liu}, Z.},
  \bibinfo{author}{v.~d. {Maaten}, L.}, \bibinfo{author}{{Weinberger}, K.Q.},
  \bibinfo{year}{2017}.
\newblock \bibinfo{title}{Densely connected convolutional networks}, in:
  \bibinfo{booktitle}{2017 IEEE Conference on Computer Vision and Pattern
  Recognition (CVPR)}, pp. \bibinfo{pages}{2261--2269}.
\bibitem[{Ioffe and Szegedyet(2015)}]{Sergey}
\bibinfo{author}{Ioffe, S.}, \bibinfo{author}{Szegedyet, C.},
  \bibinfo{year}{2015}.
\newblock \bibinfo{title}{{Batch Normalization:} accelerating deep network
  training by reducing internal covariate shift}.
\newblock \bibinfo{journal}{arXiv:1502.03167} .
\bibitem[{Jalalian et~al.(2017)Jalalian, Mashohor, Mahmud, Karasfi, Saripan and
  Ramli}]{cad2}
\bibinfo{author}{Jalalian, A.}, \bibinfo{author}{Mashohor, S.},
  \bibinfo{author}{Mahmud, R.}, \bibinfo{author}{Karasfi, B.},
  \bibinfo{author}{Saripan, M.I.B.}, \bibinfo{author}{Ramli, A.R.B.},
  \bibinfo{year}{2017}.
\newblock \bibinfo{title}{Foundation and methodologies in computer-aided
  diagnosis systems for breast cancer detection}.
\newblock \bibinfo{journal}{EXCLI Journal} \bibinfo{volume}{16},
  \bibinfo{pages}{113--137}.
\bibitem[{Kaiser et~al.(2017)Kaiser, Gomez, Shazeer, Vaswani, Parmar, Jones and
  Uszkoreit}]{Kaiser}
\bibinfo{author}{Kaiser, L.}, \bibinfo{author}{Gomez, A.N.},
  \bibinfo{author}{Shazeer, N.}, \bibinfo{author}{Vaswani, A.},
  \bibinfo{author}{Parmar, N.}, \bibinfo{author}{Jones, L.},
  \bibinfo{author}{Uszkoreit, J.}, \bibinfo{year}{2017}.
\newblock \bibinfo{title}{One model to learn them all}.
\newblock \bibinfo{journal}{CoRR} \bibinfo{volume}{abs/1706.05137}.
\bibitem[{Korotkov and Garcia(2012)}]{r28}
\bibinfo{author}{Korotkov, K.}, \bibinfo{author}{Garcia, R.},
  \bibinfo{year}{2012}.
\newblock \bibinfo{title}{Computerized analysis of pigmented skin lesions: a
  review}.
\newblock \bibinfo{journal}{Artificial intelligence in medicine}
  \bibinfo{volume}{56}, \bibinfo{pages}{69--90}.
\bibitem[{Li and Shen(2018)}]{Shen}
\bibinfo{author}{Li, Y.}, \bibinfo{author}{Shen, L.}, \bibinfo{year}{2018}.
\newblock \bibinfo{title}{Skin lesion analysis towards melanoma detection using
  deep learning network}.
\newblock \bibinfo{journal}{Sensors} \bibinfo{volume}{11}.
\bibitem[{Lin et~al.(2014)Lin, Chen and Yan}]{min}
\bibinfo{author}{Lin, M.}, \bibinfo{author}{Chen, Q.}, \bibinfo{author}{Yan,
  S.}, \bibinfo{year}{2014}.
\newblock \bibinfo{title}{Network in network}.
\newblock \bibinfo{journal}{{International Conference on Learning
  Representations (ICLR)}} .
\bibitem[{Long et~al.(2015)Long, Shelhamer and Darrell}]{long}
\bibinfo{author}{Long, J.}, \bibinfo{author}{Shelhamer, E.},
  \bibinfo{author}{Darrell, T.}, \bibinfo{year}{2015}.
\newblock \bibinfo{title}{Fully convolutional networks for semantic
  segmentation}, in: \bibinfo{booktitle}{2015 IEEE Conference on Computer
  Vision and Pattern Recognition (CVPR)}, pp. \bibinfo{pages}{3431--3440}.
\bibitem[{Melli et~al.(2006)Melli, Costantino and Cucchiara}]{r11}
\bibinfo{author}{Melli, R.}, \bibinfo{author}{Costantino, C.},
  \bibinfo{author}{Cucchiara, R.}, \bibinfo{year}{2006}.
\newblock \bibinfo{title}{Comparison of color clustering algorithms for
  segmentation of dermatological images}, in: \bibinfo{booktitle}{Medical
  Imaging 2006: Image Processing}, \bibinfo{organization}{International Society
  for Optics and Photonics}. p. \bibinfo{pages}{61443S}.
\bibitem[{Mendon{\c{c}}a et~al.(2013)Mendon{\c{c}}a, Ferreira, Marques, Marcal
  and Rozeira}]{ph2data}
\bibinfo{author}{Mendon{\c{c}}a, T.}, \bibinfo{author}{Ferreira, P.M.},
  \bibinfo{author}{Marques, J.S.}, \bibinfo{author}{Marcal, A.R.S.},
  \bibinfo{author}{Rozeira, J.}, \bibinfo{year}{2013}.
\newblock \bibinfo{title}{Ph 2-a dermoscopic image database for research and
  benchmarking}, in: \bibinfo{booktitle}{2013 35th annual international
  conference of the IEEE engineering in medicine and biology society (EMBC)},
  \bibinfo{organization}{IEEE}. pp. \bibinfo{pages}{5437--5440}.
\bibitem[{Mishraa and Celebi(2016)}]{Stoecker}
\bibinfo{author}{Mishraa, N.K.}, \bibinfo{author}{Celebi, M.E.},
  \bibinfo{year}{2016}.
\newblock \bibinfo{title}{An overview of melanoma detection in dermoscopy
  images using image processing and machine learning}.
\newblock \bibinfo{journal}{arXiv:1601.07843} .
\bibitem[{Odena et~al.(2016)Odena, Dumoulin and Olah}]{odena}
\bibinfo{author}{Odena, A.}, \bibinfo{author}{Dumoulin, V.},
  \bibinfo{author}{Olah, C.}, \bibinfo{year}{2016}.
\newblock \bibinfo{title}{Deconvolution and checkerboard artifacts},
  \bibinfo{publisher}{Distill}.
\bibitem[{Oliveira et~al.(2016)Oliveira, Filho, Z, Papa, Pereira and
  Tavares}]{Roberta}
\bibinfo{author}{Oliveira, R.B.}, \bibinfo{author}{Filho, M.E.},
  \bibinfo{author}{Z, M.}, \bibinfo{author}{Papa, J.P.},
  \bibinfo{author}{Pereira, A.S.}, \bibinfo{author}{Tavares, J.M.},
  \bibinfo{year}{2016}.
\newblock \bibinfo{title}{Computational methods for the image segmentation of
  pigmented skin lesions: A review}.
\newblock \bibinfo{journal}{Computer Methods and Programs in Biomedicine}
  \bibinfo{volume}{131}, \bibinfo{pages}{127--141}.
\bibitem[{Rasmus et~al.(2015)Rasmus, Valpola, Honkala, Berglund and
  Raiko}]{Ladder}
\bibinfo{author}{Rasmus, A.}, \bibinfo{author}{Valpola, H.},
  \bibinfo{author}{Honkala, M.}, \bibinfo{author}{Berglund, M.},
  \bibinfo{author}{Raiko, T.}, \bibinfo{year}{2015}.
\newblock \bibinfo{title}{Semi-supervised learning with ladder networks}.
\newblock \bibinfo{journal}{28th International Conference on Neural Information
  Processing Systems} \bibinfo{volume}{2}, \bibinfo{pages}{3546--3554}.
\bibitem[{Ronneberger et~al.(2015)Ronneberger, Fischer and Brox}]{olaf}
\bibinfo{author}{Ronneberger, O.}, \bibinfo{author}{Fischer, P.},
  \bibinfo{author}{Brox, T.}, \bibinfo{year}{2015}.
\newblock \bibinfo{title}{U-net: Convolutional networks for biomedical image
  segmentation}.
\newblock \bibinfo{journal}{International Conference on Medical Image Computing
  and Computer-Assisted Intervention} \bibinfo{volume}{9351},
  \bibinfo{pages}{234--241}.
\bibitem[{Salimi et~al.(2018)Salimi, Bozorgtabar, Schmid-Saugeon, Ekenel and
  Thiran}]{Saleh}
\bibinfo{author}{Salimi, S.B.}, \bibinfo{author}{Bozorgtabar, S.},
  \bibinfo{author}{Schmid-Saugeon, P.}, \bibinfo{author}{Ekenel, H.K.},
  \bibinfo{author}{Thiran, J.}, \bibinfo{year}{2018}.
\newblock \bibinfo{title}{{DermoNet:} densely linked convolutional neural
  network for efficient skin lesion segmentation}.
\newblock \bibinfo{journal}{EPFL scientific publications} .
\bibitem[{Sarker et~al.(2018)Sarker, Rashwan, Akram, Banu, Saleh, Singh,
  Chowdhury, Abdulwahab, Romani and Radeva}]{r18}
\bibinfo{author}{Sarker, M.M.K.}, \bibinfo{author}{Rashwan, H.A.},
  \bibinfo{author}{Akram, F.}, \bibinfo{author}{Banu, S.F.},
  \bibinfo{author}{Saleh, A.}, \bibinfo{author}{Singh, V.K.},
  \bibinfo{author}{Chowdhury, F.H.}, \bibinfo{author}{Abdulwahab, S.},
  \bibinfo{author}{Romani, S.}, \bibinfo{author}{Radeva, P.},
  \bibinfo{year}{2018}.
\newblock \bibinfo{title}{{SLSdeep:} skin lesion segmentation based on dilated
  residual and pyramid pooling networks}, in: \bibinfo{booktitle}{International
  Conference on Medical Image Computing and Computer-Assisted Intervention},
  \bibinfo{organization}{Springer}. pp. \bibinfo{pages}{21--29}.
\bibitem[{Smith and MacNeil(2011)}]{typeImage}
\bibinfo{author}{Smith, L.}, \bibinfo{author}{MacNeil, S.},
  \bibinfo{year}{2011}.
\newblock \bibinfo{title}{State of the art in non‐invasive imaging of
  cutaneous melanoma}.
\newblock \bibinfo{journal}{Skin Res. Technol.} \bibinfo{volume}{17},
  \bibinfo{pages}{257--269}.
\bibitem[{Tang et~al.(2018)Tang, Yang, Yuan and Zhan}]{Yujiao}
\bibinfo{author}{Tang, Y.}, \bibinfo{author}{Yang, F.}, \bibinfo{author}{Yuan,
  S.}, \bibinfo{author}{Zhan, C.A.}, \bibinfo{year}{2018}.
\newblock \bibinfo{title}{A multi-stage framework with context information
  fusion structure for skin lesion segmentation}.
\newblock \bibinfo{journal}{CoRR} \bibinfo{volume}{abs/1810.07075}.
\bibitem[{Wen(2017)}]{Wen}
\bibinfo{author}{Wen, H.}, \bibinfo{year}{2017}.
\newblock \bibinfo{title}{Ii-fcn for skin lesion analysis towards melanoma
  detection}.
\newblock \bibinfo{journal}{arXiv:1702.08699} .
\bibitem[{Yamashita et~al.(2018)Yamashita, Nishio, Do and Togashi}]{Rikiya}
\bibinfo{author}{Yamashita, R.}, \bibinfo{author}{Nishio, M.},
  \bibinfo{author}{Do, R.K.G.}, \bibinfo{author}{Togashi, K.},
  \bibinfo{year}{2018}.
\newblock \bibinfo{title}{Convolutional neural networks: an overview and
  application in radiology}.
\newblock \bibinfo{journal}{Insights into Imaging} \bibinfo{volume}{9},
  \bibinfo{pages}{611--629}.
\bibitem[{Yuan(2017)}]{yuan2017automatic}
\bibinfo{author}{Yuan, Y.}, \bibinfo{year}{2017}.
\newblock \bibinfo{title}{Automatic skin lesion segmentation with fully
  convolutional-deconvolutional networks}.
\newblock \bibinfo{journal}{arXiv:1703.05165} .
\bibitem[{Yuan and Lo(2019)}]{jaccard}
\bibinfo{author}{Yuan, Y.}, \bibinfo{author}{Lo, Y.}, \bibinfo{year}{2019}.
\newblock \bibinfo{title}{Improving dermoscopic image segmentation with
  enhanced convolutional-deconvolutional networks}.
\newblock \bibinfo{journal}{IEEE Journal of Biomedical and Health Informatics}
  \bibinfo{volume}{23}, \bibinfo{pages}{519--526}.
\bibitem[{Zeiler(2012)}]{adadelta}
\bibinfo{author}{Zeiler, M.D.}, \bibinfo{year}{2012}.
\newblock \bibinfo{title}{Adadelta: an adaptive learning rate method}.
\newblock \bibinfo{journal}{CoRR} \bibinfo{volume}{abs/1212.5701}.
\bibitem[{Zhou et~al.(2013)Zhou, Li, Schaefer, Celebi and Millera}]{Zhou}
\bibinfo{author}{Zhou, H.}, \bibinfo{author}{Li, X.},
  \bibinfo{author}{Schaefer, G.}, \bibinfo{author}{Celebi, M.E.},
  \bibinfo{author}{Millera, P.}, \bibinfo{year}{2013}.
\newblock \bibinfo{title}{Mean shift based gradient vector flow for image
  segmentation}.
\newblock \bibinfo{journal}{Computer Vision and Image Understanding}
  \bibinfo{volume}{117}, \bibinfo{pages}{1004--1016}.

\end{thebibliography}

\clearpage
\onecolumn
\appendix
\section{Prediction Samples}
\label{appendix1}

The predicted segmentation masks (Pred) using the proposed network overlaid on the ground truth (GT) where TP, FP, and FN are depicted in green, yellow, and red respectively. 
The first 6 rows depict the results for the challenging images presented in Fig. \ref{fig:TrainingImages} whereas the last 3 rows are the worst predictions obtained using our network.

\begin{table}[!ht]  
\centering 
\scriptsize
{
\begin{tabular}{c c||| c c}
\textbf{Original Image}  & \textbf{Pred and GT Overlay} & \textbf{Original Image}  & \textbf{Pred and GT Overlay}  \\

\includegraphics[width=3cm,height=1.7cm]{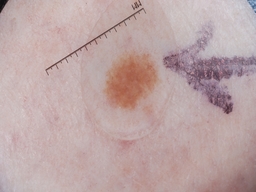}   &
\includegraphics[width=3cm,height=1.7cm]{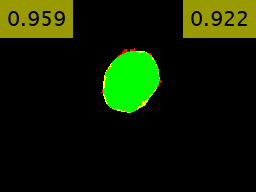}  &

\includegraphics[width=3cm,height=1.7cm]{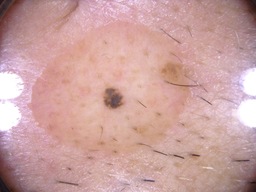}   &
\includegraphics[width=3cm,height=1.7cm]{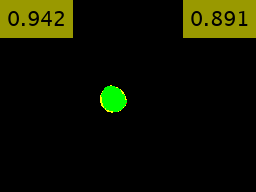} \\ \hline

\includegraphics[width=3cm,height=1.7cm]{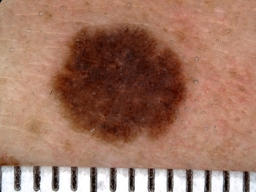}   &
\includegraphics[width=3cm,height=1.7cm]{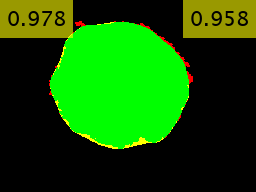}  &

\includegraphics[width=3cm,height=1.7cm]{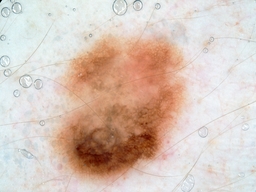}   &
\includegraphics[width=3cm,height=1.7cm]{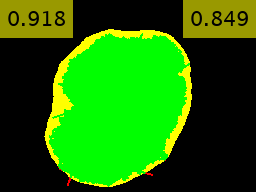} \\ \hline

\includegraphics[width=3cm,height=1.7cm]{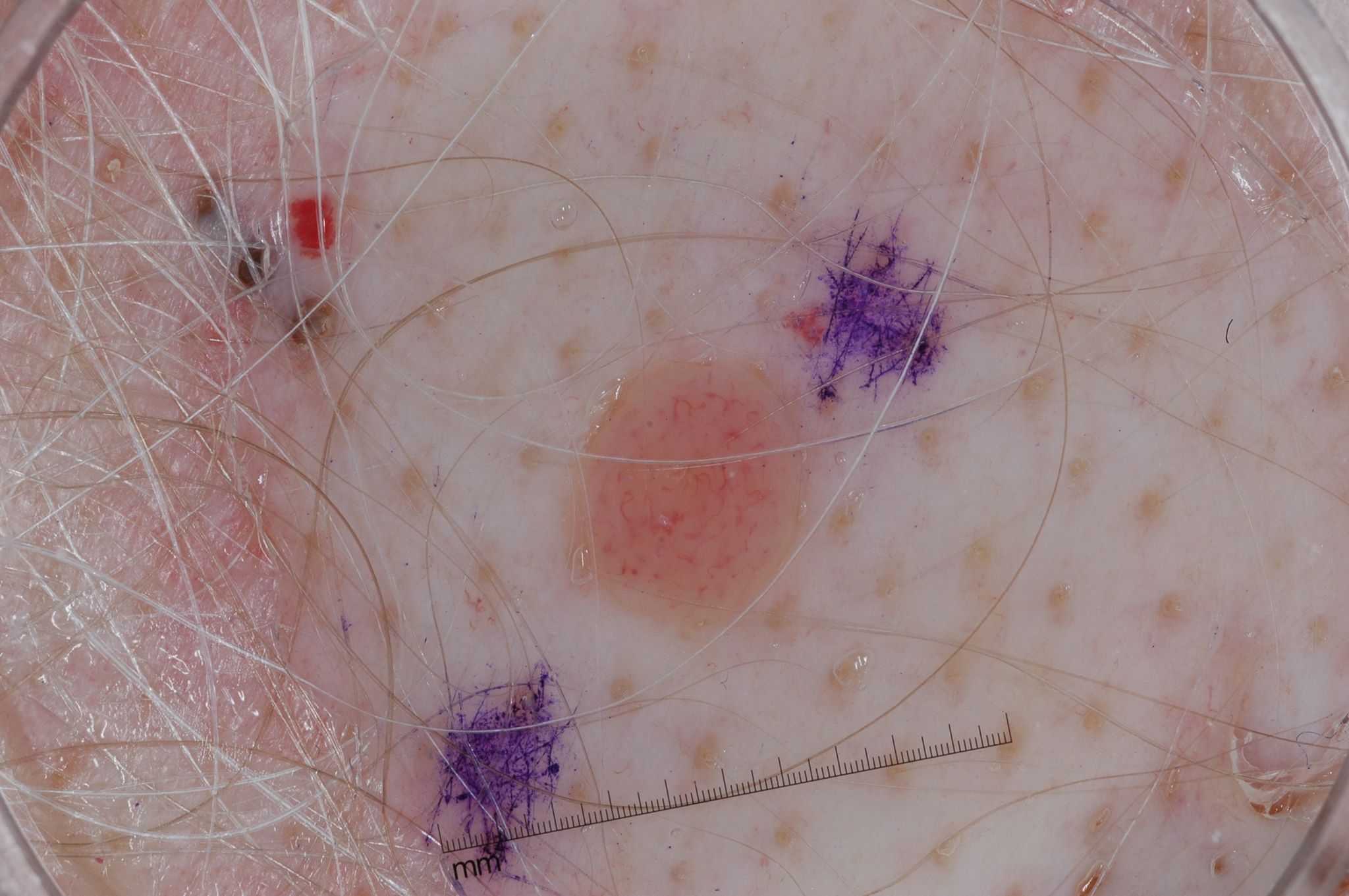}   &
\includegraphics[width=3cm,height=1.7cm]{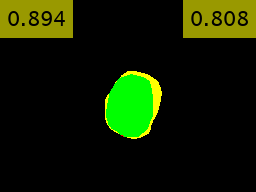}  &

\includegraphics[width=3cm,height=1.7cm]{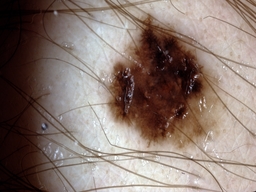}   &
\includegraphics[width=3cm,height=1.7cm]{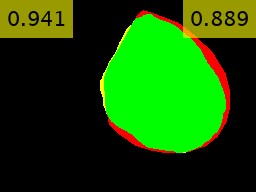} \\ \hline

\includegraphics[width=3cm,height=1.7cm]{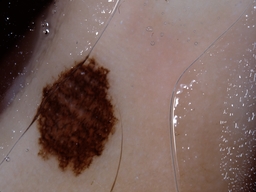}   &
\includegraphics[width=3cm,height=1.7cm]{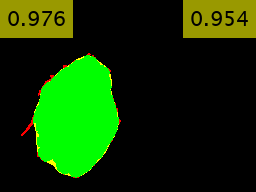}  &

\includegraphics[width=3cm,height=1.7cm]{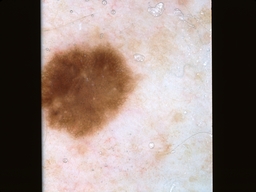}   &
\includegraphics[width=3cm,height=1.7cm]{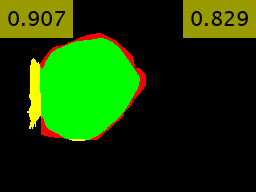} \\ \hline

\includegraphics[width=3cm,height=1.7cm]{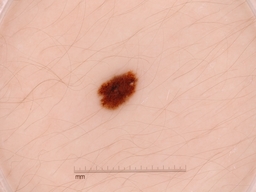}   &
\includegraphics[width=3cm,height=1.7cm]{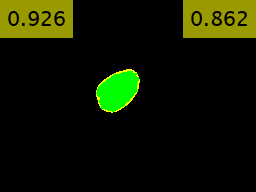}  &

\includegraphics[width=3cm,height=1.7cm]{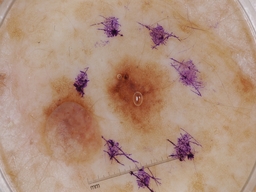}   &
\includegraphics[width=3cm,height=1.7cm]{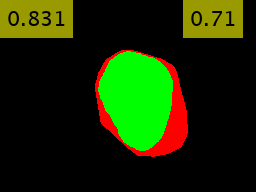} \\ \hline

\includegraphics[width=3cm,height=1.7cm]{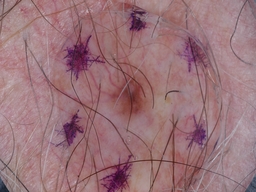}   &
\includegraphics[width=3cm,height=1.7cm]{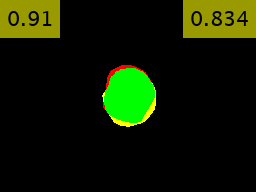}  &

\includegraphics[width=3cm,height=1.7cm]{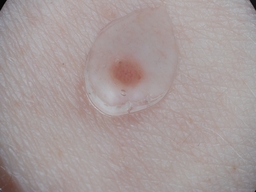}   &
\includegraphics[width=3cm,height=1.7cm]{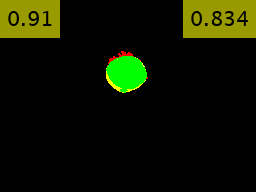} \\ \hline

\includegraphics[width=3cm,height=1.7cm]{Appendix/org/ISIC_0015353.jpg}   &
\includegraphics[width=3cm,height=1.7cm]{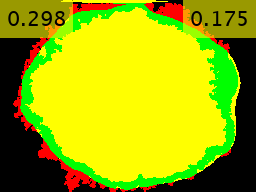}  &

\includegraphics[width=3cm,height=1.7cm]{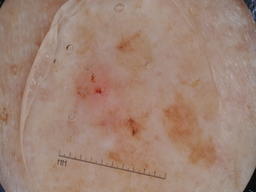}   &
\includegraphics[width=3cm,height=1.7cm]{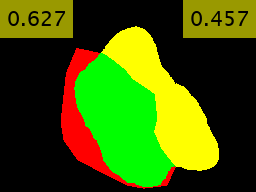} \\ \hline

\includegraphics[width=3cm,height=1.7cm]{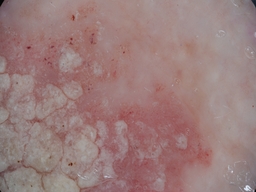}   &
\includegraphics[width=3cm,height=1.7cm]{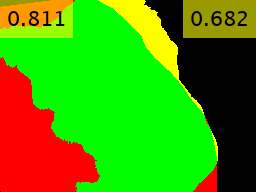}  &

\includegraphics[width=3cm,height=1.7cm]{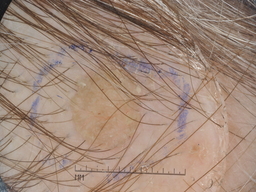}   &
\includegraphics[width=3cm,height=1.7cm]{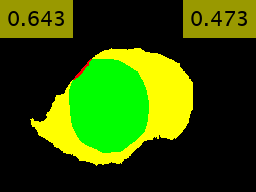} \\ \hline

\includegraphics[width=3cm,height=1.7cm]{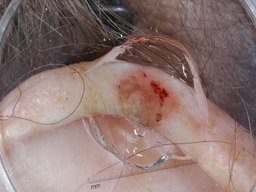}   &
\includegraphics[width=3cm,height=1.7cm]{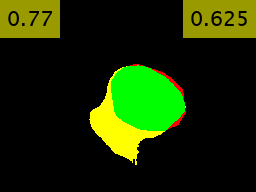}  &

\includegraphics[width=3cm,height=1.7cm]{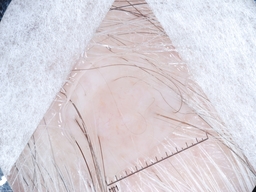}   &
\includegraphics[width=3cm,height=1.7cm]{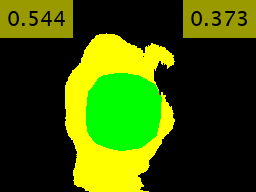} \\ \hline

\end{tabular}

}  
\end{table}  

\end{document}